\documentclass[conference]{IEEEtran}

\IEEEoverridecommandlockouts

\newtheorem{theorem}{Theorem}

\usepackage{amssymb}

\usepackage{cite}

\ifCLASSINFOpdf

\else

\fi

\usepackage{amsmath}

\usepackage{algorithm}
\usepackage{algorithmicx}
\usepackage{algpseudocode}

\usepackage{array}

\usepackage{flushend}

\usepackage{makecell}
\usepackage{multirow}

\usepackage{graphicx}
\usepackage{subfigure}

\usepackage{times}
\usepackage{array}
\newlength\savedwidth

\newlength\savewidth
\newcommand\shline{\noalign{\global\savewidth\arrayrulewidth
                            \global\arrayrulewidth 1.5pt}%
                   \hline
                   \noalign{\global\arrayrulewidth\savewidth}}

\begin{document}

\title{To Bond or not to Bond: An Optimal Channel Allocation Algorithm For Flexible Dynamic Channel Bonding in WLANs}

\author{Caihong Kai $^{\dagger}$ Yuting Liang $^{\dagger}$ Tianyu Huang $^{\dagger}$ and Xu Chen $^{\ast}$\\
$^{\dagger}$School of Computer Science and Information Engineering, Hefei University of Technology, Hefei, China\\
$^{\ast}$School of Data and Computer Science, Sun Yat-sen University, China\\
Email:chkai@hfut.edu.cn, \{yutingliang, tyHuang\}@mail.hfut.edu.cn, chenxu35@mail.sysu.edu.cn
\thanks{This work was partially supported by NSFC project (No.61571178 and 61202459.).}}

\maketitle

\begin{abstract}
IEEE 802.11 has evolved from 802.11a/b/g/n to 802.11ac to meet rapidly increasing data rate requirements in WLANs. One important technique adopted in 802.11ac is the channel bonding (CB) scheme that combines multiple 20MHz channels for a single transmission in 5GHz band. In order to effectively access channel after a series of contention operations, 802.11ac specifies two different CB operations: Dynamic Channel Bonding (DCB) and Static Channel Bonding (SCB). This paper proposes an optimal channel allocation algorithm to achieve maximal throughputs in DCB WLANs. Specifically, we first adopt a continuous-time Markov Chain (CTMC) model to analyze the equilibrium throughputs. Based on the throughput analysis, we then construct an integer nonlinear programming (INLP) model with the target of maximizing system throughput. By solving the INLP model, we then propose an optimal channel allocation algorithm based on the Branch-and-Bound Method (BBM). It turns out that the maximal throughput performance can be achieved under the channel allocation scheme with the least overlapped channels among WLANs. Simulations show that the proposed algorithm can achieve the maximal system throughput under various network settings. We believe that our analysis on the optimal channel allocation schemes brings new insights into the design and optimization of future WLANs, especially for those adopting channel bonding technique.
\end{abstract}

\begin{IEEEkeywords}
802.11ac; Dynamic Channel Bonding; WLANs; CSMA Protocol, Channel Allocation
\end{IEEEkeywords}

\IEEEpeerreviewmaketitle

\section{Introduction}
It is known that the channel bonding (CB) technique has been used in wireless networks to boost data rates. The adoption of the CB technique in WLANs was first introduced in the IEEE 802.11n amendment \cite{09std}, where two basic 20MHz channels can be aggregated to obtain a 40MHz channel. To support high-speed applications, the IEEE802.11ac amendment \cite{14std} further extends the allowable bandwidth in a single transmission from 40MHz to 80MHz and even 160MHz. The design target of 802.11ac is to offer very high throughput (VHT) while keep backward compatibility with the legacy 802.11 specifications \cite{perahia2011gigabit}. However, the usage of wider channels also makes the channel contention between the neighboring WLANs more complicated, in which the contending node is allowed to dynamically select its transmission channels based on the instantaneous spectrum occupancy status just at the beginning of the transmission. Such a CB technique is usually referred to as Dynamic Channel Bonding (DCB) \cite{park2011ieee}. This paper makes an attempt to analyze the interactions and dependencies under DCB and seek for the optimal channel allocation strategy to maximize the aggregate throughputs in DCB WLANs.

\begin{figure}
  \centering
  \includegraphics[width=3.5in]{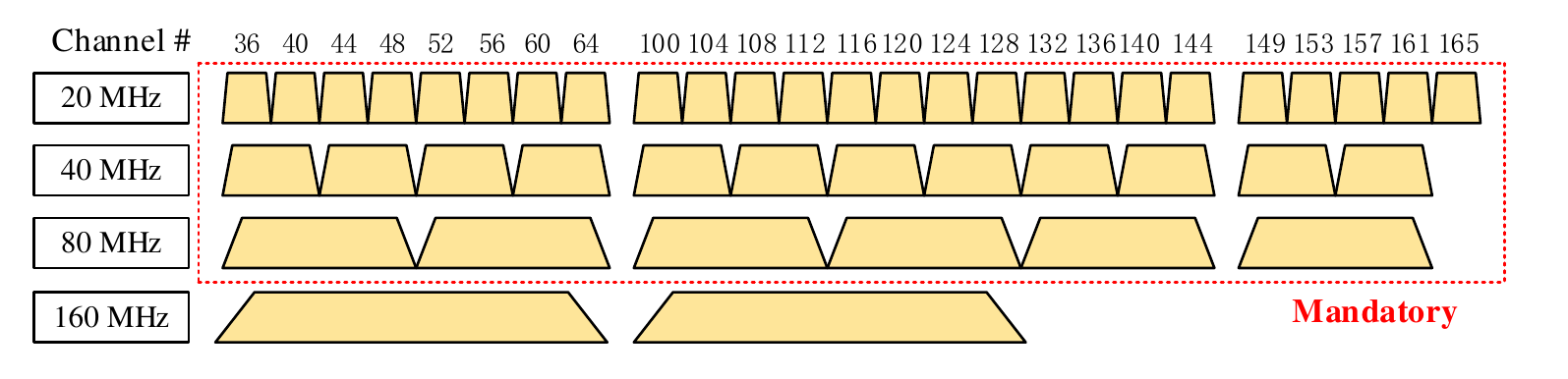}\\
  \caption{Channel Allocation Map in 5 GHz Band of 802.11ac [2].}\label{1}
\end{figure}

There have been several studies on the performance of 802.11ac networks \cite{park2011ieee,gong2011channel,deek2014intelligent,srikanth2016performance,bellalta2016interactions,faridi2016analysis}. By simulations, refs. [4, 5] showed that the channel bonding technique can provide significant throughput gains. An experimental evaluation on different network parameters that affect the performance of the CB in IEEE 802.11 WLANs was presented in [6]. From the perspective of performance analysis, [7] proposed an analytical model based on a decoupling approximation to evaluate the performance of an IEEE 802.11ac WLAN with dynamic bandwidth channel access. Ref. [8] constructed a Continuous-Time Markov Chain (CTMC) to analyze the network performance under the Static Channel Bonding (SCB). Later, ref. [9] extends the CTMC model to analyze the interactions between neighboring WLANs operating under DCB, in which all WLANs are ``all-inclusive" in the sense that all WLANs can sense each other. However,  although several research efforts have been made to analyze the performance of DCB networks, none of them have been devoted to investigate the impacts of different channel allocation schemes (such as the number of basic channels and the location of the primary channel of each WLAN) on the system performance in DCB networks and this paper attempts to fill this gap.

It is known that WLANs adopts the CSMA/CA protocol for multiple user access at the MAC layer, whose main components are carrier sensing and random backoff to alleviate packet collisions. In DCB networks, the maximum number of basic channels that can be used by a WLAN and the selection of the primary channel are two important parameters that affect the interactions and dependencies among WLANs and lead to different system performance. Observing this, we investigate the optimal channel allocation algorithms to achieve maximal throughputs in DCB WLANs. More specifically, we first adopt the CTMC model proposed in [9] to analyze the throughput performance under different channel allocations. Importantly, we prove that for “all-inclusive” DCB networks, the optimal throughput performance is achieved under one of the channel allocation scheme with the least overlapped channels among WLANs. Based on this understanding, we then construct an integer nonlinear programming (INLP) model with the target of maximizing system throughput. By solving the INIP model, we then proposed an optimal channel allocation algorithm based on the BBM to seek for the optimal channel allocation scheme that maximizes the aggregate throughputs of all WLANs. Simulation results validate that the proposed channel allocation algorithm can achieve the maximal network throughput and maintain good fairness among WLANs for ``all-inclusive" DCB networks.

We believe that our analysis on channel allocations of ``all-inclusive" DCB networks provides new insights into the 802.11ac networks and moves a signigicant step towards the optimization of the DCB networks. For example, by theoretical analysis we show that the too much overlapped channels among WLANs could in fact decrease the aggregate throughputs under current 802.11ac parameter settings\footnote{This result was also observed by experimental evaluation in [6]}, thus it is nontrivial to determine how to bond basic channels in WLANs.  Moreover, the proposed channel allocation scheme can achieve optimal throughput performance and is suitable for engineering implementation in practical WLANs.

The remainder of the paper is organized as follows. Section II introduces the background on wide bandwidth operation defined in 802.11ac as well as the channel allocation algorithms in WLANs. Section III describes the channel allocation problem of DCB networks, then introduces the throughput computation using the CTMC model and performs numerical analysis to find out the effect of different channel allocation schemes on the network throughput performance. Section IV presents the throughput analysis under different channel allocation schemes and the constructed INLP model, then we propose a channel allocation algorithm based on BBM to solve the INLP problem in Section V. Simulations are shown in Section VI. Finally Section VII concludes this paper.

\section{Background}
\subsection{Channelization and Channel Contention Defined in 802.11ac}
802.11ac allows WLANs to use multiple non-overlapping channels [2], [3] in a single transmission. As shown in Fig.1, two adjacent 20MHz channels can form a 40MHz channel, and two adjacent 40MHz channels can form an 80MHz channel. A 160MHz channel can be formed by two adjacent or separated 80MHz channels. We call a 20MHz channel as a basic channel. To support this expanded channelization, each node uses control fields in the beacon to indicate its bandwidth and the selected primary channel [2].

Under CB, a wider bandwidth channel is composed of a primary channel and one or more secondary channels. Each node in the network uses the basic distributed coordination function (DCF) to compete for channel occupancies only on the primary channel [2], [3]. When a node has packets to transmit, it first senses its primary channel. Once the primary channel has been sensed idle for a DCF inter-frame space (DIFS) duration, the node starts the backoff procedure by selecting a random value of the backoff counter. The node then starts decreasing the backoff timer linearly with time while sensing the primary channel idle. If the primary channel is sensed busy during the backoff process, the backoff timer is frozen with the remaining time recorded. Upon the primary channel is sensed idle for a DIFS time again, the backoff process resumes with the recorded remaining time.

Different from the case of single-channel WLANs, before the timer expires, the node has to sense its secondary channels for a point coordination function (PCF) inter-frame space (PIFS) period. When the time expires, the node has two options to determine to transmit on which channels: i) under SCB, as shown in Fig. 2(a), only when all the channels (including both the primary and the non-primary channels) are idle, the node starts transmitting using the whole assigned channel. Otherwise, it will initializes a new backoff procedure; ii) under DCB, as shown in Fig. 2(b), even though some of the non-primary channels may be busy during the PIFS, the node begins to transmit using the primary channel and the idle non-primary channels that are adjacent to the primary channel, without initializing a new backoff process. It is known that DCB has much better performance than SCB [4], and thus this paper considers the DCB WLAN.

\begin{figure}

  \begin{minipage}[b]{0.2\textwidth}
  \centering
  \includegraphics[width=3.5in]{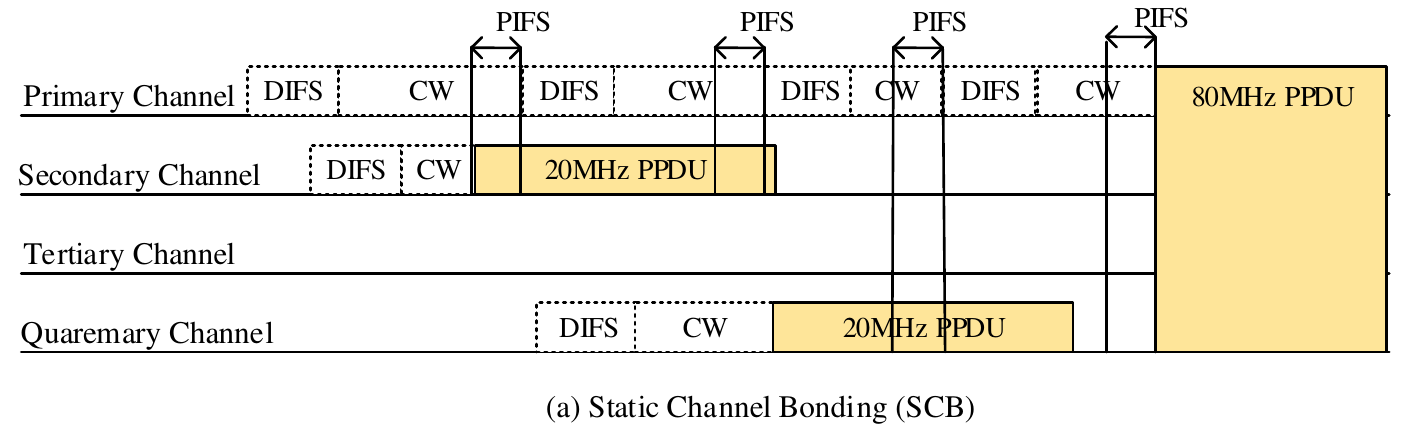}\\
  \includegraphics[width=3.5in]{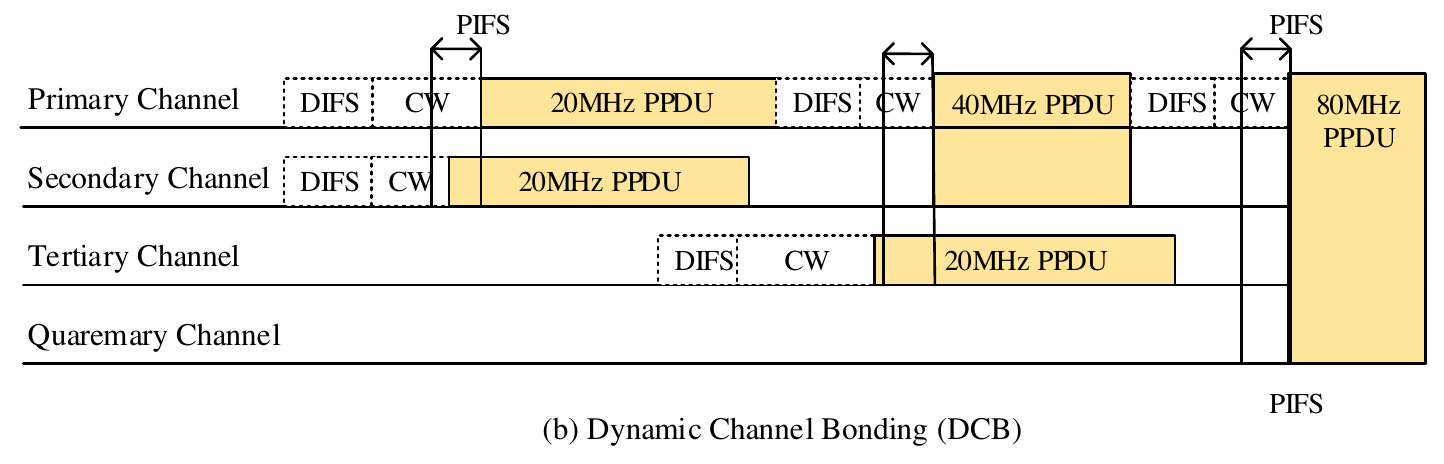}
  \end{minipage}
  \caption{Two Types of Channel Bonding Schemes in 802.11ac[4].}\label{2}
\end{figure}

\subsection{Channel Allocation Algorithms in WLANs}
Channel allocation algorithms in WLANs have attracted much interest from the research community \cite{mahonen2004automatic,mishra2005weighted,chieochan2010channel,mengual2013channel,kamiya2015joint}. It is commonly cast as graph-coloring where an edge corresponds to interference between two cells, and the set of available colors corresponds to the set of channels. Because graph-coloring is NP-hard for general graphs, heuristics are used to solve it [10]-[14]. However, most of prior works do not take into account the challenges brought by the CB technique. The authors in \cite{arslan2010auto} proposed an algorithm simultaneously consider the channel center frequency and channel bandwidth to increase throughputs gains per-AP. Ref.\cite{gong2014distributed} developed an analytical model to estimate the network throughput under client interferences and proposed a distributed channel assignment algorithm. Another decentralized algorithm was proposed in \cite{herzen2013distributed} to select both the channel center frequency and the channel width by sensing the interference that is caused by the other neighboring WLANs. Ref. \cite{wang2016managing} proposed a practical distributed protocol-compatible channel bonding scheme. However, none of [15]-[18] considered the DCB operation in channel competition.

There are also some investigations on the channel allocation schemes in SCB networks. For example, \cite{jang2015channel} analyzed the hidden terminal problem and proposed a channel allocation algorithm considering the primary channel selection with a given channel width in SCB networks. Moreover, ref. [8] proposed a solution based on the water-filling concept to find the sub-optimal allocation in SCB networks. However, considering the different channel access options between SCB and DCB, the channel allocation in DCB networks deserves a more careful investigation, and this paper makes such an attempt. It is not difficult to see that the maximum number of channels of each WLAN and the selection of the primary channel are two important parameters that affect the interactions and dependencies among WLANs and lead to different system performance, and a good channel allocation algorithm should specify the settings of both parameters above.

\section{System Model}
This section presents our network model and gives an introduction to the throughput computation of DCB WLANs based on the CTMC model proposed in [9]. What’s more, through the numerical analysis, we can roughly see the effect of different channel allocation schemes on the network throughput performance, that is, the less overlapped channels among WLANs in the network, the better throughput performance it can achieve.

\subsection{Network Model and Problem Formulation}
Similar to [9], we consider a DCB network with $N$ WLANs, in which all WLANs are within the carrier-sensing range of each other (i.e., ``all-inclusive" DCB networks). We assume a WLAN-centric model that all nodes in each WLAN are close to each other . Furthermore, all WLANs are assumed to be saturated, and we also assume when a node in WLANs initializes a transmission, the channel allocated to it will be used until the end of this transmission. That is, the node cannot switch between different channel allocations during a single transmission.

Let $K$ be the number of available basic channels (i.e., number of 20MHz channels), $C$ be all possible combinations of these basic channels (specified by the 802.11ac standard as shown in Fig. 1), and $F$ be the set of all possible channel allocations of the entire network. Define a feasible channel allocation $\textbf{\emph{f}} = \left[ {{f_1},{f_2},{f_3}, \cdots ,{f_N}} \right]$ as the vector indicating the channels assigned to all WLANs in the network, where ${f_i} \in C$ denotes the channels assigned to a single WLAN, $WLA{N_i}$. For example, ${f_i} = \left\{ {1,\tilde 2,3,4} \right\}$ denotes that $WLA{N_i}$ is allocated basic channels from 1 to 4 and the assigned primary channel is channel 2. Let ${k_i}$ be the number of contiguous basic channels assigned to $WLA{N_i}$, and the bandwidth assigned to $WLA{N_i}$ is then $B{W_i} = 20{k_i}{\rm{ MHz}}$.

Let $T{h_i}\left( \textbf{\emph{f}} \right)$ be the equilibrium throughput of $WLA{N_i}$ under the channel allocation scheme, $\textbf{\emph{f}}$, whose computation will be presented in Part B. The problem of finding the optimal channel allocation, $\textbf{\emph{f*}}$, to maximize the system throughput, can be formulated as the following optimization problem:
\begin{equation}
\begin{array}{l}
{\rm{OPT1:}} \qquad \qquad {\max _\textbf{\emph{f}}}{\rm{ }}\sum\limits_{i = 1}^N {T{h_i}\left( \textbf{\emph{f}} \right)} \\
{\rm{}} \qquad \qquad \qquad \quad s.t. \ {\rm{  }}\textbf{\emph{f}} \in F,{f_i} \in C
\end{array}
\end{equation}

For a DCB network with $N$ WLANs and $K$ available basic channels, let $\left| C \right|$ be the number of possible combinations of basic channels, then the number of all possible channel allocations is $\left| F \right| = {\left| C \right|^N}$, which grows exponentially with $N$. That is, the searching for the optimal channel allocation in the feasible region is of high complexity.

\subsection{Throughput Computation using the CTMC Model}
We next briefly introduce the CTMC model proposed in [9] for throughput computation of a DCB network under a specific channel allocation scheme $\textbf{\emph{f}} = \left[ {{f_1},{f_2},{f_3}, \cdots ,{f_N}} \right]$. For more details, interested readers are referred to [9]. In this paper, we pay more attention to find out the optimal channel allocation scheme that maximizes the aggregate throughput of the DCB network, which is absent in [9].

Under DCB, even two or more WLANs are assigned overlapped basic channels, they could still be transmitting simultaneously as long as the channels they use in this specific transmission do not overlap. Also, in a DCB network a WLAN may occupy different numbers of basic channels in different transmissions, which is different from the case of SCB as studied in [8]. The selected channels for transmission of a node in $WLA{N_i}$ based on the status of the basic channels in ${f_i}$, which are sensed just before the backoff timer reaches zero. Thus, we define a feasible network state as a set of channels on which WLANs are transmitting simultaneously, and we define the state space, $S$, as the set composed of all feasible states\footnote{Since the interactions of channel competitions under SCB and DCB are different, the feasible states and state transitions are different even under the same network settings.}.

\begin{figure}
  \centering
  \includegraphics[width=3.0in]{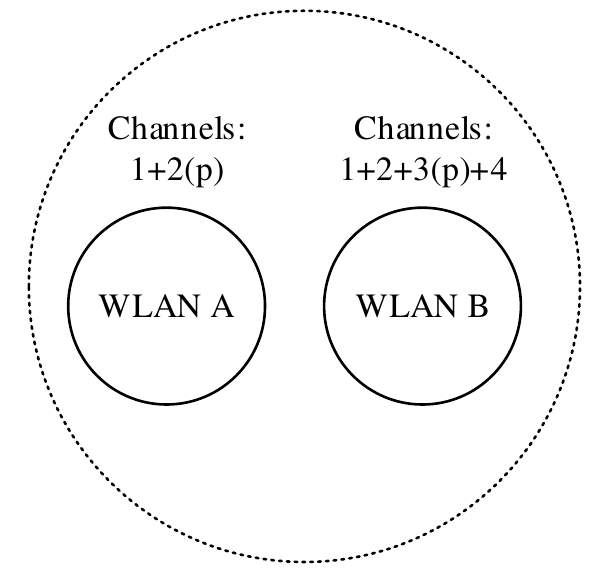}\\
  \caption{An Illustrating Example to Explain the CTMC Model}\label{3}
\end{figure}

We use an illustrating example of Fig.3, where two neighboring WLANs are within the carrier-sensing range of each other, and the channel allocation scheme is $\textbf{\emph{f}}:{f_A} = \left\{ {1,\tilde 2} \right\},{f_B} = \left\{ {1,2,\tilde 3,4} \right\}$, to demonstrate the computation. The set of feasible network states is $S = \left\{ {\emptyset ,A_2^1,B_4^1,A_2^1B_2^3,B_2^3} \right\}$, where $\emptyset $ is the state in which none of the WLANs is transmitting, $A_2^1$ and $B_4^1$ is the network state in which only WLAN A or B is transmitting and using channels $\{ 1,2\} $ or $\{ 1,2,3,4\} $ respectively. The top number of $_ \bullet ^ \bullet $ is the first selected channel of $WLA{N_i}$ and the bottom number is the total number of basic channels used by $WLA{N_i}$ for the current transmission. The transmission channels selected under DCB is always the largest contiguous subset of these available channels that contains the primary channel. Similarly, $A_2^1B_2^3$ is the network state in which the two WLANs are simultaneously transmitting, using channels $\{ 1,2\} $ and $\{ 3,4\} $ respectively.

We assume that the backoff timer at each node is in continuous time and has an average duration of $E\left[ {{B_i}} \right]$ seconds, where ${B_i}$ is the backoff duration of $WLA{N_i}$. We define the attempt rate of a node in $WLA{N_i}$ as ${\lambda _i} = E{\left[ {{B_i}} \right]^{ - 1}}$ when it has a packet waiting for transmission. The transmission duration is denoted by ${T_i}\left( {{{k'}_i},{\gamma _i},{L_i}} \right)$, which is determined by the number of basic channels that are bound together in this particular transmission, ${k'_i}$, the Signal-to-Noise Ratio(SNR) observed by all packet transmissions inside $WLA{N_i}$, ${\gamma _i}$, and the payload size ${L_i}$. Therefore, the packet departure rate from a node in $WLA{N_i}$ is ${\mu _i} = E{\left[ {{T_i}\left( {{{k'}_i},{\gamma _i},{L_i}} \right)} \right]^{ - 1}}$. Then the transition rates between two network states, $s,s' \in S$, are
\begin{equation}
q\left( {s,s'} \right) = \left\{ \begin{array}{l}
{\lambda _i}{\rm{     }} \quad if{\rm{ }}s' = s \cup \left\{ {WLA{N_i}} \right\}\\
{\mu _{i{\rm{          }}}} \quad if{\rm{ }}s' = s\backslash \left\{ {WLA{N_i}} \right\}\\
0{\rm{        }} \qquad \qquad \ otherwise.
\end{array} \right.
\end{equation}

Define the activity ratio of $WLA{N_i}$ as the ratio of the mean packet transmission duration to the mean backoff time. That is,
\begin{equation}
  {\rho _i}\left( {{{k'}_i}} \right){\rm{ = }}\frac{{E\left[ {T\left( {{{k'}_i},{\gamma _i},{L_i}} \right)} \right]}}{{E\left[ {{B_i}} \right]}} = \frac{{{\lambda _i}}}{{{\mu _i}}}
\end{equation}

It is worthwhile to note that in a DCB network a WLAN may occupy different numbers of basic channels in different transmissions(i.e., ${k'_i}$ of $WLA{N_i}$ in different transmissions are different), which is quite different from the case of SCB as studied in [8].

Let $s\left( t \right) \in S$ denote the network state at time $t$. If we further assume the backoff and transmission durations are exponentially distributed, $s{\left( t \right)_{t \ge 0}}$ is a continuous-time Markov process on the state space $S$. This Markov process is aperiodic, irreducible and thus positive recurrent, since the state space $S$ is finite. A steady-state solution to the CTMC always exists, and we denote the stationary probability distribution as ${\left\{ {{\pi _s}} \right\}_{s \in S}}$.

The steady-state probabilities of the CTMC can be computed by solving the general balance equations, yields
\begin{equation}
  {\pi _s} = {\pi _\emptyset }\prod\limits_{i \in s} {{\rho _i}\left( {{{k'}_i}} \right)}
\end{equation}
where ${\pi _\emptyset }$ denotes the steady-state probability of the network state where none of the WLANs is activating and $i \in s$ means a node in $WLA{N_i}$ is transmitting packets in network state $s$. Together with the normalizing condition $\sum {_{s \in S}{\pi _s}}  = 1$, yields
\begin{equation}
  {\pi _\emptyset } = \frac{1}{{\sum {_{s \in S}} \prod {_{i \in s}{\rho _i}\left( {{{k'}_i}} \right)} }}
\end{equation}
and
\begin{equation}
  {\pi _s} = \frac{{\prod {_{i \in s}{\rho _i}\left( {{{k'}_i}} \right)} }}{{\sum {_{s \in S}} \prod {_{i \in s}{\rho _i}\left( {{{k'}_i}} \right)} }},{\rm{  }}s \in S
\end{equation}

Since the process $s{\left( t \right)_{t \ge 0}}$ is irreducible and positive recurrent on $S$, it follows from the classical Markov chains results that ${\pi _s}$ is equal to the long-run fraction of the time that the system spends on state $s$.

The stationary distribution of the illustrating example shown in Fig.3 is: ${\pi _{A_2^1}} = {\rho _A}\left( 2 \right){\pi _\emptyset }$, ${\pi _{B_4^1}} = {\rho _B}\left( 4 \right){\pi _\emptyset }$, ${\pi _{A_2^1B_2^3}} = {\rho _A}\left( 2 \right){\rho _B}\left( 2 \right){\pi _\emptyset }$, ${\pi _{B_2^3}} = {\rho _B}\left( 2 \right){\pi _\emptyset }$ with ${\pi _\emptyset }{\rm{ = }}{\left( {1{\rm{ + }}{\rho _A}\left( 2 \right) + {\rho _B}\left( 2 \right){\rm{ + }}{\rho _B}\left( 4 \right) + {\rho _A}\left( 2 \right){\rho _B}\left( 2 \right)} \right)^{ - 1}}$, where ${\rho _A}\left( 2 \right)$ means the activity ratio of   using two basic channels for current transmission.

It is worth mentioning that it has been proven theoretically that in SCB and DCB networks with continuous-backoff time, the stationary distribution of the Markov chain is insensitive to the distributions of both the backoff and the transmission time [8][9]. Indeed, even if the backoff and transmission time are not exponentially distributed, we can still use the continuous time Markov chain to compute the network throughput.

Based on the steady-state probabilities, we can compute the throughputs of WLANs. The throughput of $WLA{N_i}$, in bits per second, is then given by
\begin{equation}
T{h_i} = {L_i}\left( {\sum {_{s \in S,i \in s}{\mu _i}{\pi _s}} } \right)\left( {1 - {p_e}} \right)
\end{equation}
where ${p_e}$ is the packet error probability. In our model we assume that the channel is ideal and ${p_e} = 0$.

It is important to note that any change in channel allocation scheme, $\emph{\textbf{f}}$, results in a different state space, $S$, as well as the different transitions among them. As can be seen in (5)(6), the normalization constant ${\pi _\emptyset }$ and the stationary distribution ${\left\{ {{\pi _s}} \right\}_{s \in S}}$ depend on the state space $S$ and the state transitions among them. Thus, different channel allocation schemes will lead to different throughput performances. It’s crucial to find an optimal channel allocation scheme that can lead to a maximal aggregate throughput.

\subsection{Numerical Analysis}
In order to find out the effect of different channel allocation schemes on the network throughput performance, we analyze the throughput performance of a simple DCB network, which is composed of four neighboring WLANs, under different channel allocation schemes, as shown in Fig. 4, where each block represents the assigned basic channel and each block with diagonals represents the assigned primary channel. The number of available basic channels is set to $K{\rm{ = }}4$. Scenario 1 represents the case that all WLANs are allocated the same set of basic channels, we name it as ``totally-overlapped". The case that all WLANs are allocated a set of non-overlapped channels is showed in scenario 2, named as ``non-overlapped". Scenario 3 represent a random-allocation case that there are partial overlapped channels among WLANs, named as ``partially-overlapped". The position of primary channel of each WLAN in the network is not overlap in the above three scenarios, that is, all WLANs in the network have different positions of primary channel. Therefore, in scenario 4, the set of basic channels assigned to each WLAN is the same as scenario 3, but with the positions of primary channel overlapping condition, named as ``partially-primary-overlapped".

\begin{figure}
  \centering
  \includegraphics[width=3.5in]{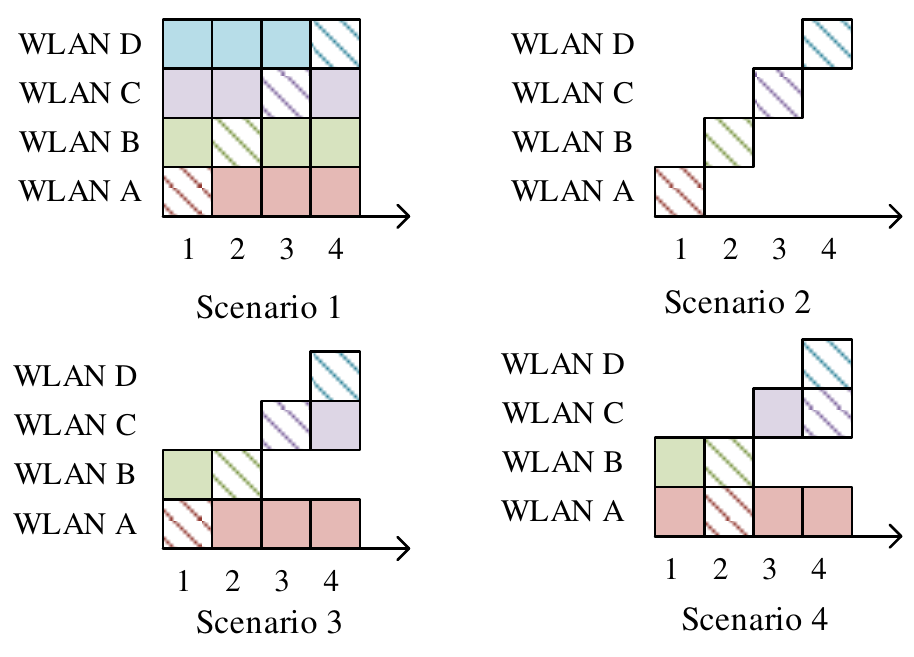}\\
  \caption{Four Channel Allocation Schemes}\label{4}
\end{figure}

Using the aforementioned throughput computation method in Part B, we can get the throughputs of four scenarios, which can be denoted as $T{h_{to}}$, $T{h_{no}}$, $T{h_{po}}$, $T{h_{ppo}}$ for scenario ``totally-overlapped", ``non-overlapped", ``partially-overlapped", and ``partially-primary-overlapped" respectively:
\begin{equation}
  \begin{array}{l}
T{h_{to}} = \frac{{4\lambda L}}{{1 + 4\rho \left( 4 \right)}}\\
T{h_{no}} = \frac{{4\lambda L}}{{1 + \rho \left( 1 \right)}}\\
T{h_{po}} = \frac{{\left( {6 + 8\rho \left( 2 \right) + 6\rho \left( 1 \right) + 2{\rho ^2}\left( 1 \right) + 4\rho \left( 1 \right)\rho \left( 2 \right)} \right)\lambda L}}{{1 + \rho \left( 4 \right) + 3\rho \left( 2 \right) + 2\rho \left( 1 \right) + 2{\rho ^2}\left( 2 \right) + 4\rho \left( 1 \right)\rho \left( 2 \right) + {\rho ^2}\left( 1 \right) + 2{\rho ^2}\left( 1 \right)\rho \left( 2 \right)}}\\
T{h_{ppo}} = \frac{{\left( {5 + 6\rho \left( 2 \right) + 2\rho \left( 1 \right)} \right)\lambda L}}{{1 + \rho \left( 4 \right) + 3\rho \left( 2 \right) + \rho \left( 1 \right) + 2{\rho ^2}\left( 2 \right) + 2\rho \left( 1 \right)\rho \left( 2 \right)}}
\end{array}
\end{equation}

In addition, we assume all WLANs have same size of the transmitted packet, $L$ and have same attempt rate, $\lambda $. We denote the normalized throughput of four scenarios shown in (8) as $Th'$ by remove $\lambda L$ in each equation, and with the parameters presented in Section VI, we can get
\begin{equation}
  \begin{array}{l}
\rho \left( 1 \right) = \frac{{E\left[ {T\left( 1 \right)} \right]}}{{E\left[ B \right]}} = 170.2778\\
\rho \left( 2 \right) = \frac{{E\left[ {T\left( 2 \right)} \right]}}{{E\left[ B \right]}} = 92.0833\\
\rho \left( 4 \right) = \frac{{E\left[ {T\left( 4 \right)} \right]}}{{E\left[ B \right]}} = 64.4444
\end{array}
\end{equation}
and
\begin{equation}
  \begin{array}{l}
T{h'_{to}}{\rm{ = 0}}{\rm{.0155}}\\
T{h'_{no}}{\rm{ = 0}}{\rm{.0234}}\\
T{h'_{po}}{\rm{ = 0}}{\rm{.0225}}\\
T{h'_{ppo}}{\rm{ = 0}}{\rm{.0184}}
\end{array}
\end{equation}

Obviously, $T{h'_{no}} > T{h'_{po}} > T{h'_{ppo}} > T{h'_{to}}$ can hold. A rough thought come into mind that the less overlapped channels among WLANs in the network, the better throughput performance it can achieve.

We also write a simulator to describe the network operations under DCB, as in [9], we use a continuous time bachoff and assume the propagation delay is zero, which results in a zero collision probability. The achieved throughput of each WLAN is presented in Fig.5 to give us an insight of the interactions among WLANs. The point denotes the output of CTMC model and the bar denotes the mean value average over 1000 simulations.

\begin{figure}
\begin{minipage}[t]{0.48\linewidth}
  \centerline{\includegraphics[width=4.0cm]{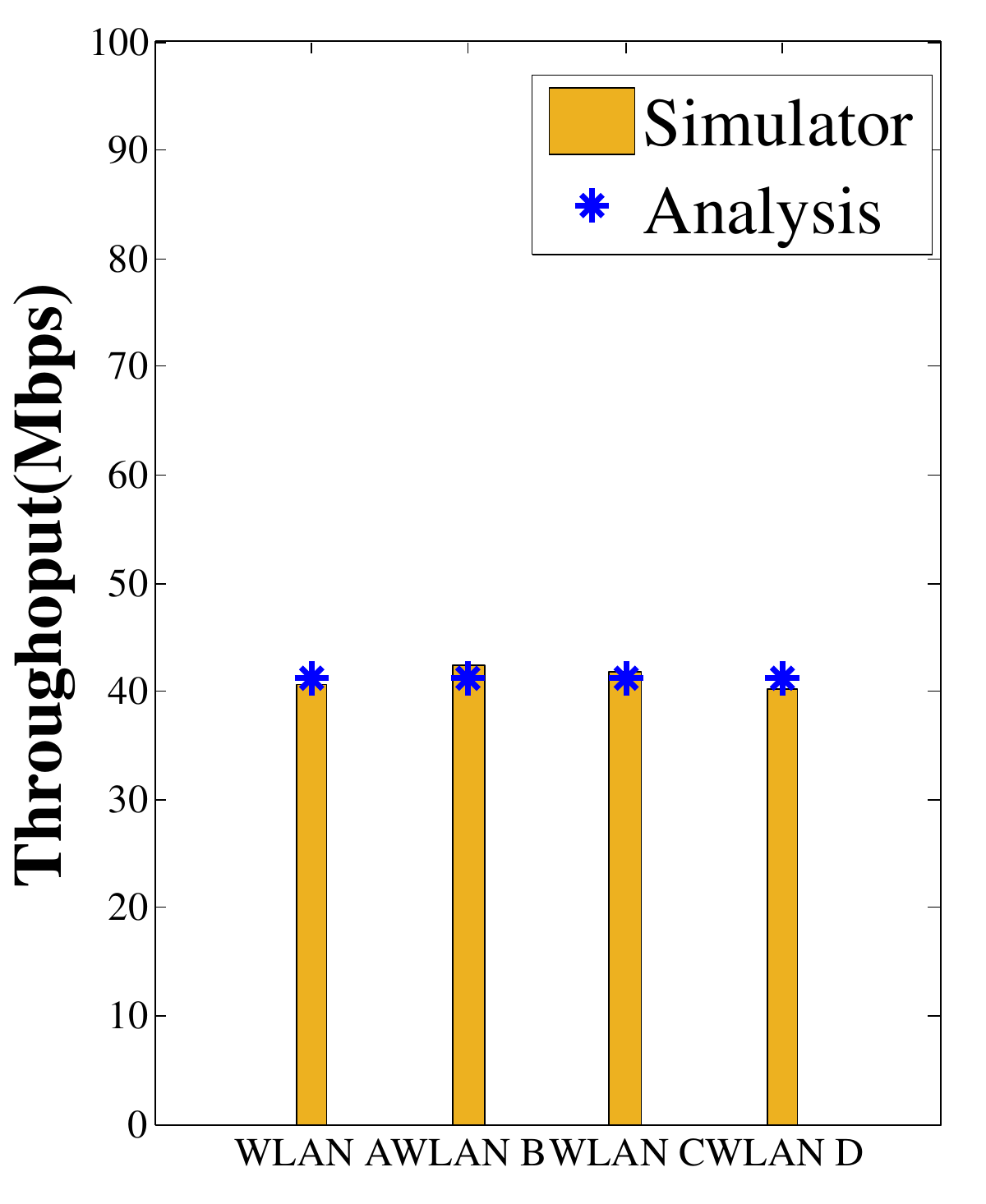}}
  \centerline{(a) Scenario 1}
\end{minipage}
\hfill
\begin{minipage}[t]{0.48\linewidth}
  \centerline{\includegraphics[width=4.0cm]{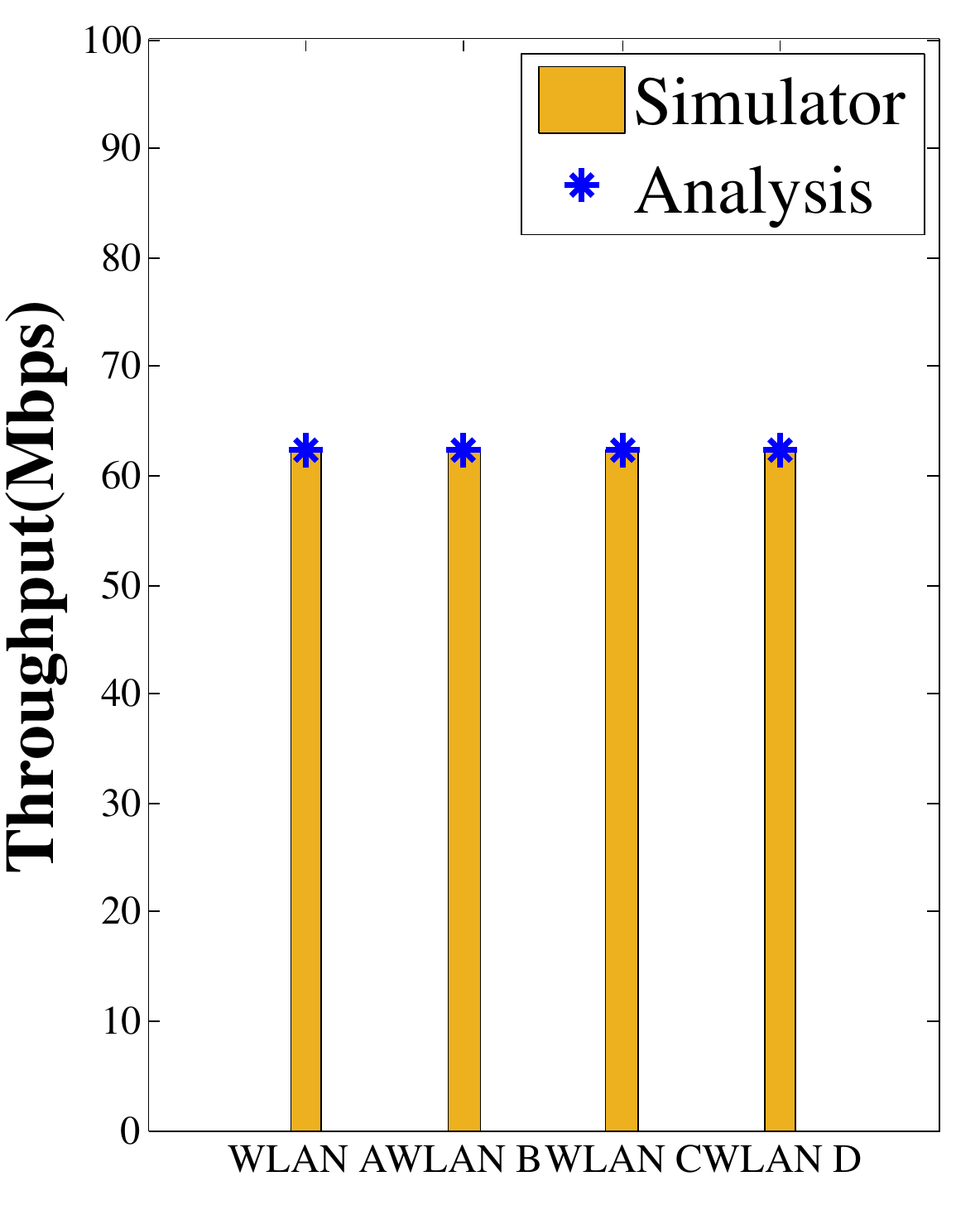}}
  \centerline{(b) Scenario 2}
\end{minipage}
\vfill
\begin{minipage}[t]{0.48\linewidth}
  \centerline{\includegraphics[width=4.0cm]{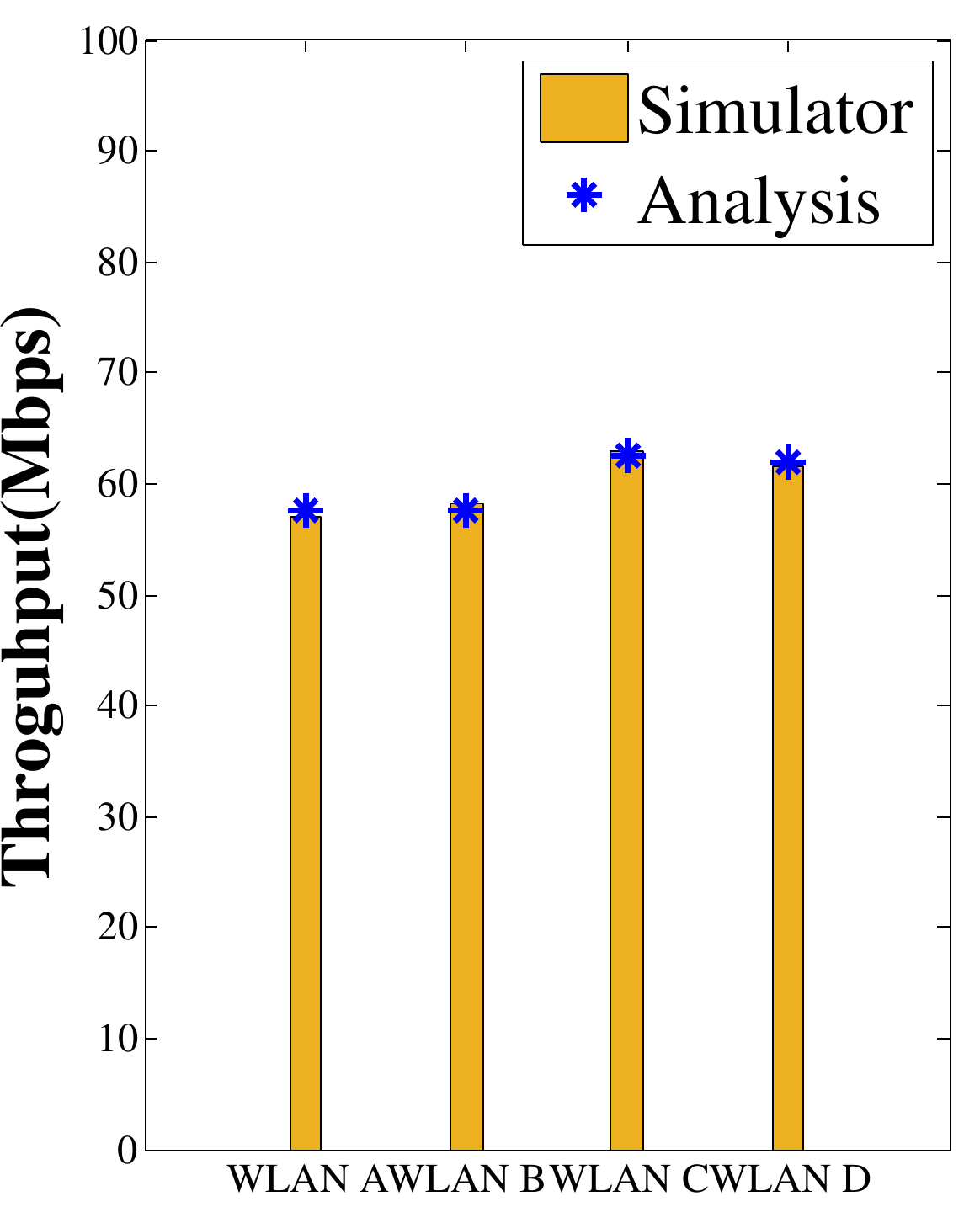}}
  \centerline{(c) Scenario 3}
\end{minipage}
\hfill
\begin{minipage}[t]{0.48\linewidth}
  \centerline{\includegraphics[width=4.0cm]{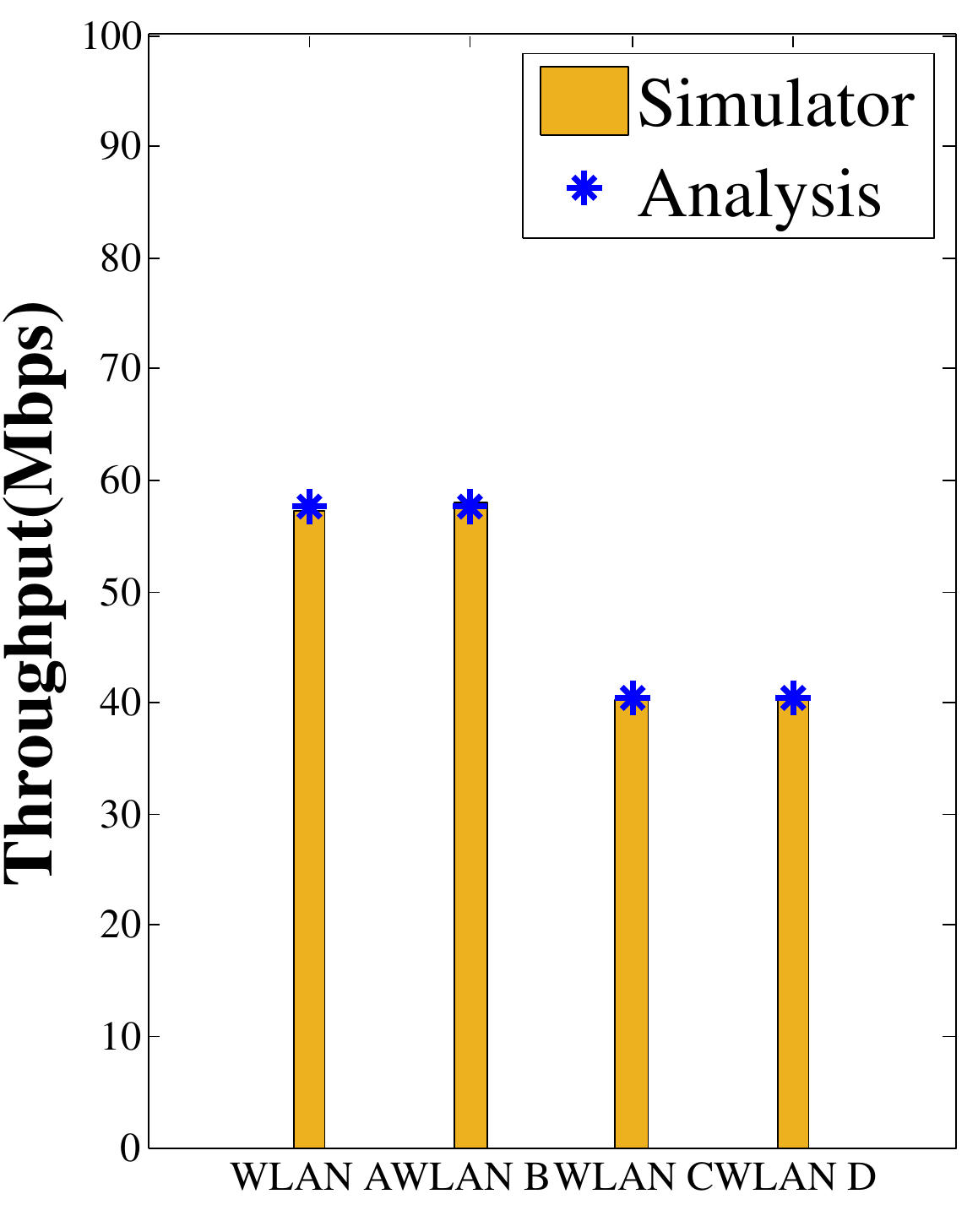}}
  \centerline{(d) Scenario 4}
\end{minipage}
\caption{Achieved Throughput of Each WLAN}
\label{5}
\end{figure}

From Fig.5, on the one hand, we can see the analytical and simulated results match well, which validates the correctness of the CTMC model. On the other hand, we can get two interesting findings. In scenario 1, the set of feasible network states is $S = \left\{ {\emptyset ,A_4^1,B_4^1,C_4^1,D_4^1} \right\}$, four WLANs are all assigned the same set of basic channels, consequently they compete with all of the others for the channels, and get the same transmission probability for all WLANs in the long term, which results in the same throughput for all of them. In scenario 2, the set of feasible network states is $S{\rm{ = }}\left\{ {\emptyset ,A_1^1,B_1^1,C_1^1,D_1^1,A_1^1B_1^1,A_1^1C_1^1,A_1^1D_1^1,B_1^1C_1^1,B_1^1D_1^1} \right.$ $\left. {,C_1^1D_1^1,A_1^1B_1^1C_1^1,A_1^1B_1^1D_1^1,A_1^1C_1^1D_1^1,B_1^1C_1^1D_1^1,A_1^1B_1^1C_1^1D_1^1} \right\}$, all WLANs are allocated non-overlapped channels, which means they are completely independent of each other, each WLAN can be treated as an independent system. In scenario 3, the set of feasible network states is $S = \left\{ {\emptyset ,A_4^1,B_2^1,C_2^3,D_1^4,A_2^1,C_1^3,} \right.A_2^1C_2^3,A_2^1D_1^4,B_2^1C_2^3,B_2^1D_1^4$, $\left. {C_1^3D_1^4,A_2^1C_1^3D_1^4,B_2^1C_1^3D_1^4} \right\}$ , $WLA{N_A}$ has been allocated total basic channels, thus it has a partial set of channels that is same with the set of channels allocated to all the other three WLANs respectively. In this case, although $WLA{N_A}$ has to compete with all the other three WLANs for the channels, it can also simultaneously transmit with $WLA{N_C}$ or $WLA{N_D}$ due to the channel access scheme in use, which is DCB as we described in background. Still, $WLA{N_A}$ can't transmit with $WLA{N_B}$ at the same time, because if the primary channel of $WLA{N_A}$ is free (the backoff timer in channel 1 reaches zero), and $WLA{N_A}$ has packets to transmit, then whenever $WLA{N_A}$ finds channel 2 available, it will integrate channel 1 and 2 as one channel for transmission, or channel 1 to 4 if they are all available, $WLA{N_B}$ cannot be transmitting at the same time. Thus $WLA{N_A}$ can get the same throughput as $WLA{N_B}$. A comparison of the throughputs of scenarios 1 to 3 indicates that the channel allocation scheme with less number of overlapped basic channels, has a better throughput performance, which is our first interesting finding. In scenario 4, the sets of basic channels assigned to the four WLANs are the same as scenario 3, the only difference is that they have different allocations of the primary channel. In this case, the set of feasible network states is $S = \left\{ {\emptyset ,A_4^1,B_2^1,C_2^3,D_1^4,A_2^1,A_2^1C_2^3,A_2^1D_1^4,B_2^1C_2^3,B_2^1D_1^4} \right\}$, $WLA{N_A}$
has the same primary channel with $WLA{N_B}$, and $WLA{N_C}$ has the same primary channel with $WLA{N_D}$, thus, they can’t transmit together for the WLANs who have the same position of primary channel. A comparison of the throughputs of scenario 3 and 4 indicates that the channel allocation scheme with non-overlapped primary channel can get a better throughput performance, which is our second interesting finding.

Given the problem we formulated and the numerical analysis, we notice that the channel bonding technical is not always helpful under current IEEE 802.11ac parameter settings, thus, it is nontrivial to determine how to bond basic channels in WLANs. We then move forward to analyze the throughput performance of different channel allocation schemes and make efforts to get the optimal solution to achieve the maximal system throughput for DCB networks.

\section{Performance Analysis of Channel Allocation Schemes in DCB Nerworks}
This section first carefully examines the throughput performance under different channel allocation schemes and builds up an integer nonlinear programming (INLP) model with the target of maximizing system throughput. By theoretical analysis we figure out that the optimal throughput performance is achieved under the channel allocation scheme with the least overlapped channels among WLANs. This observation is the basis of our proposed channel allocation algorithm to be presented in Section V.

Consider a DCB network composed of $N$ WLANs that are all in the carrier-sensing range of each other. We also assume that there are four basic channels available (note that we will remove this assumption in Section VII, and our analysis still holds.) in the DCB network, which is labeled from channel (1) to (4). Fig. 6 shows the channel index and all possible combinations of basic channels.

\begin{figure}
  \centering
  \includegraphics[width=3.5in]{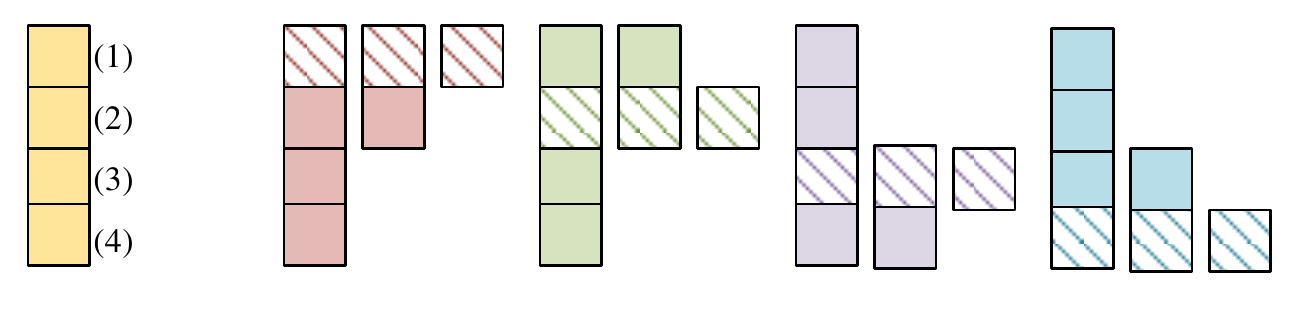}\\
  \caption{Channel Index and All Possible Combinations of Basic Channels}\label{6}
\end{figure}

Without loss of generality, we assume that the nodes in all WLANs transmit packets of a fixed size $L$, have a backoff process of equal average durations $E\left[ B \right] = {\lambda ^{ - 1}}$, and use the same modulation and coding rate regardless of the number of basic channels selected in the transmission. Therefore, if two WLANs use the same number of basic channels for a transmission, they have equal packet transmission durations.

Under a particular channel allocation scheme, $\emph{\textbf{f}}$, we say that WLANs $i$ and $j$ do not overlap if ${f_i} \cap {f_j}{\rm{ = }}\emptyset $. For WLANs with overlapped basic channels, let ${{\cal O}_{\cal X}}$ be a set of WLANs with overlapped basic channels ${\cal X}$, where ${\cal X}$ is the intersection of basic channels of all WLANs in set ${{\cal O}_{\cal X}}$, i.e., ${f_i} \cap {f_j} = {\cal X},\forall i,j \in {{\cal O}_{\cal X}},i \ne j$. We define the number of overlapped channels of $WLA{N_i}$ as ${O_i}$, indicating that $WLA{N_i}$ has ${O_i}$ channels that interact with other WLANs in the network. It is easy to see that ${O_i} \le K,\forall i$. Three situations listed as following based on the overlapped set   belongs in to calculate ${O_i}$:
\begin{itemize}
\item  If $WLA{N_i}$ doesn’t belong to any of the overlapped set, it means that the set of basic channels allocated to $WLA{N_i}$ doesn’t interact with any other WLANs in the network, then ${O_i} = 0$;
\item  If $WLA{N_i}$ only belongs to one overlapped set, i.e., $i \in {{\cal O}_{\cal X}}$, then ${O_i}{\rm{ = }}\left| {\cal X} \right|$, where $\left| {\cal X} \right|$ is the cardinality of set ${\cal X}$;
\item  If $WLA{N_i}$ belongs to more than one overlapped set, i.e., $i \in {{\cal O}_{{{\cal X}_1}}}{\rm{ }}and{\rm{ }}i \in {{\cal O}_{{{\cal X}_2}}}$, then ${O_i}$ is the cardinality of the union of basic channels in these overlapped sets, ${O_i} = \left| {{{\cal X}_1} \cup {{\cal X}_2}} \right|$.
\end{itemize}

Then let $O\left( \emph{\textbf{f}} \right){\rm{ = }}\max \left\{ {{O_i},\forall i} \right\}$ be the number of overlapped channels under channel allocation scheme $\emph{\textbf{f}}$. After further examination, we find that the optimal channel allocation scheme exhibits specific features, as shown in Theorem 1.

\begin{theorem}
Let ${\cal F}$ be the collection of all channel allocation schemes with the minimum $O\left( \textbf{\emph{f}} \right)$. For ``all-inclusive" DCB networks, the channel allocation scheme that achieves the maximal system throughput belongs to ${\cal F}$. That is, ${F^*} = \left\{ {{\textbf{\emph{f}}^*}:ma{x_\textbf{\emph{f}}}{\rm{ }}\sum\limits_{i = 1}^N {T{h_i}} \left( \textbf{\emph{f}}\right)} \right\}$ and we have ${\textbf{\emph{f}}^*} \in {\cal F}$.
\end{theorem}

\emph{Proof:} We separate the analysis into two parts:

\emph{1) When the number of WLANs $N$ is no more than the number of available basic channels $K$ (i.e., $N \le K$):}

We consider a DCB network with four basic channels. When $N \le K$, for each $N$, we can enumerate all possible channel allocation schemes and write out all expressions of the corresponding achieved throughputs. Then, we can compare these expressions through simple mathematical calculation to find the maximal throughput, which corresponds to the optimal channel allocation scheme.

To save space, we will not enumerate all possible channel allocation schemes for each $N$ and only present our analytical comparisons.

i) $N = 1$: There is only one WLAN in the network, it is obvious that the maximal throughput can be obtained by letting the WLAN use all the channels. Thus, the optimal channel allocation scheme is $\textbf{\emph{f}}:{f_1} = \left\{ {\tilde 1,2,3,4} \right\}$.

ii) $N = 2$: For two WLANs that are within the carrier sensing range, different channel allocation schemes can classify into three categories: ``totally-overlapped" (i.e., $\textbf{\emph{f}}:{f_1} = {f_2} = \left\{ {\tilde 1,2,3,4} \right\}$), ``partially-overlapped" (i.e., $\textbf{\emph{f}}:{f_1} = \left\{ {\tilde 1,2,3,4} \right\},{f_2} = \left\{ {\tilde 1,2} \right\}$ and ``non-overlapped" (i.e., $\textbf{\emph{f}}:{f_1} = \left\{ {\tilde 1,2} \right\},{f_2} = \left\{ {\tilde 3,4} \right\}$. In addition, we also consider the effect of the position of primary channel in each case. If two WLANs have the same primary channel, they will never transmit packets simultaneously. After careful examination, we find that the optimal channel allocation scheme is $\textbf{\emph{f}}:{f_1} = \left\{ {\tilde 1,2} \right\},{f_2} = \left\{ {\tilde 3,4} \right\}$. That is, there is no overlapped channels between two WLANs, and the primary channel of each WLAN does not overlap either.

iii) $N = 3$: In this case, although the total number of channel allocation schemes increases, we can still classify them into three categories and make comparisons. The optimal scheme is $\textbf{\emph{f}}:{f_1} = \left\{ {\tilde 1,2} \right\},{f_2} = \left\{ {\tilde 3} \right\},{f_3} = \left\{ {\tilde 4} \right\}$, and there is neither overlapped basic channels nor overlapped primary channels over these three WLANs.

iv) $N = 4$: After examination, we find that the optimal channel allocation scheme is $\textbf{\emph{f}}:{f_1} = \left\{ {\tilde 1} \right\},{f_2} = \left\{ {\tilde 2} \right\},{f_3} = \left\{ {\tilde 3} \right\},{f_4} = \left\{ {\tilde 4} \right\}$

When $N \le K$, we have ${\cal F} = \left\{ {\textbf{\emph{f}}:O\left( \textbf{\emph{f}} \right){\rm{ = 0}}} \right\}$. In the first three cases, there is more than one channel allocation scheme with $O\left( \textbf{\emph{f}} \right){\rm{ = 0}}$. Summarizing these four scenarios above, we can see that the channel allocation scheme achieves the maximum network throughput always exists in ${\cal F}$, so that Theorem 1 stands when $N \le K$.

 From Theorem 1, when $N \le K$ and the system has sufficient basic channels to afford each WLAN non-overlapped channels. A set of contiguous non-overlapped basic channels can be assigned to each WLAN individually in order to get the channel allocation scheme with minimum $O\left( \textbf{\emph{f}} \right)$, that is $O\left( \textbf{\emph{f}} \right) = 0$ . Therefore, we need to divide $K$ basic channels into $N$ groups and let a WLAN use a group of basic channels. That is, when $N \le K$, the CB technical can be used to boost aggregate throughput. With a litter abuse of notion, we let ${k_i}$ denote the number of non-overlapped channels allocated to  $WLA{N_i}$. In this case, the throughput achieved by $WLA{N_i}$ is $\frac{{\lambda L}}{{1 + \rho \left( {{k_i}} \right)}}$, where $\rho \left( {{k_i}} \right) = \frac{{E\left[ {T\left( {{k_i}} \right)} \right]}}{{E\left[ B \right]}}$ is the activity ratio of $WLA{N_i}$ using ${k_i}$ basic channels. Then OPT1 can be rewritten as the following INLP problem:
\begin{subequations}
\begin{equation}
  {\rm{OPT2:}} \   \max {\rm{ }}h\left( {{k_1}, \cdots ,{k_N}} \right) = \sum\limits_{i = 1}^N {T{h_i}}  = \sum\limits_{i = 1}^N {\frac{A}{{1 + \rho \left( {{k_i}} \right)}}}
\end{equation}
\begin{equation}
 s.t.  {\rm{  }}\sum\limits_{i = 1}^N {{k_i} \le K}
\end{equation}
\begin{equation}
\qquad \qquad {k_i}{\rm{ = }}{{\rm{2}}^j},j = 0,1,2,3
\end{equation}
\end{subequations}
where $A{\rm{ = }}\lambda L$ is a constant in our assumptions. The objective function $h\left( {{k_1}, \cdots ,{k_N}} \right)$ in (11.a) is the aggregate network throughput under the grouping scheme ${\cal K} = \left[ {{k_1}, \cdots ,{k_N}} \right]$. Eq.(11.b) means that the total number of basic channels assigned to WLANs cannot exceed $K$. Eq.(11.c) is the channel allocation constraint specified by the IEEE 802.11ac standard. Specifically, the number of the basic channels assigned to any WLAN could ${2^j}$ and $j = 0,1,2,3$. That is, the number of all basic channels cannot exceed eight (i.e., the largest bandwidth allowed by 802.11ac is 160MHz).

Note that the solution of OPT2 is an optimal grouping scheme, ${{\cal K}^ * } = \left[ {k_1^*, \cdots ,k_N^*} \right]$, and the optimal channel allocation scheme can be easily obtained by setting ${f_i} = \left\{ {1 + \sum\nolimits_{e < i} {\min \left( {k_e^ * ,8} \right)} , \cdots ,\sum\nolimits_{e \le i} {\min \left( {k_e^ * ,8} \right)} } \right\}$ ($e,i \in \left\{ {1, \cdots N} \right\}$) and setting the first channel in ${f_i}$ as the primary channel of $WLA{N_i}$.

Since $\rho \left( {{k_i}} \right)$ is a discrete function of ${k_i}$, to solve OPT2, we first use the calibration curve fitting to generate a continuous function of $\rho \left( {{k_i}} \right)$, denoted as $\rho '\left( {{k_i}} \right)$, and relax ${k_i}$ to a continuous variable. Then OPT2 can be converted to a non-integer nonlinear programming (NINLP) problem:
\begin{subequations}
\begin{equation}
  {\rm{OPT3:     }}\max {\rm{ }}h'\left( {{k_1}, \cdots ,{k_N}} \right) = \sum\limits_{i = 1}^N {T{h_i}}  = \sum\limits_{i = 1}^N {\frac{A}{{1 + \rho '\left( {{k_i}} \right)}}}
\end{equation}
\begin{equation}
s.t. \quad {\rm{  }}\sum\limits_{i = 1}^N {{k_i} \le K}
\end{equation}
\end{subequations}
where $\rho '\left( {{k_i}} \right) = \frac{b}{{{{\left( {{k_i}} \right)}^a}}}$, $a$ and $b$ are fitting parameters to guarantee the correlation coefficient between $\rho \left( {{k_i}} \right)$ and $\rho '\left( {{k_i}} \right)$ is above 0.98, and we set $a = 0.7624$, $b=168.2$.

We next comply with the standard solution of Lagrange Multiplier Approach (LMA) to solve OPT3. Introducing a Lagrange multiplier $\gamma $ ($\gamma  \ge 0$) to establish a Lagrange function $H\left( {{k_1}, \cdots ,{k_N},\gamma } \right)$ first, which is
\begin{equation}
  H\left( {{k_1}, \cdots ,{k_N},\gamma } \right){\rm{ = }}\sum\limits_{i = 1}^N {\frac{A}{{1 + \rho '\left( {{k_i}} \right)}}}  + \gamma \left( {{k_1} +  \cdots  + {k_N} - K} \right)
\end{equation}
then we take the derivatives of $H\left( {{k_1}, \cdots ,{k_N},\gamma } \right)$ for all unknown variables, and make them equal to zero to get the extreme point, which can be represented as:
\begin{equation}
  \begin{array}{l}
\frac{{\partial H}}{{\partial {k_1}}} = \frac{{Aabk_1^{ - a - 1}}}{{{{\left[ {1 + \rho '\left( {{k_1}} \right)} \right]}^2}}} + \gamma  = 0\\
{\rm{              }} \vdots \\
\frac{{\partial H}}{{\partial {k_N}}} = \frac{{Aabk_N^{ - a - 1}}}{{{{\left[ {1 + \rho '\left( {{k_N}} \right)} \right]}^2}}} + \gamma  = 0\\
\frac{{\partial H}}{{\partial \gamma }} = {k_1} +  \cdots  + {k_N} - K = 0
\end{array}
\end{equation}

By solving (14), we can get an grouping scheme, ${\cal K} = \left[ {\bar k, \cdots ,\bar k} \right]$, where $\bar k = K/N$. Note $\bar k$ can be interpreted as the mean number of non-overlapped channels that allocated to each WLAN, which might be a non-integer.

As for (12.b), the constraint is linear, consequently, we need to prove that the objective function is a concave function. To this end, we need prove the odd-order partial derivation of ${\rm \textbf{H}}$ is always negative and the even-order derivation of ${\rm \textbf{H}}$ is always positive. The Hessian matrix of (12.a) is
\[{\rm \textbf{H}} = \left\{ {\begin{array}{*{20}{c}}
{\frac{{\partial {h^2}}}{{{\partial ^2}{k_1}}}}&{\frac{{\partial {h^2}}}{{\partial {k_1}\partial {k_2}}}}& \cdots &{\frac{{\partial {h^2}}}{{\partial {k_1}\partial {k_N}}}}\\
{\frac{{\partial {h^2}}}{{\partial {k_2}\partial {k_1}}}}&{\frac{{\partial {h^2}}}{{{\partial ^2}{k_2}}}}& \cdots &{\frac{{\partial {h^2}}}{{\partial {k_2}\partial {k_N}}}}\\
 \vdots & \vdots & \ddots & \vdots \\
{\frac{{\partial {h^2}}}{{\partial {k_N}\partial {k_1}}}}&{\frac{{\partial {h^2}}}{{\partial {k_N}\partial {k_2}}}}& \cdots &{\frac{{\partial {h^2}}}{{{\partial ^2}{k_N}}}}
\end{array}} \right\}\]
the elements of ${\rm \textbf{H}}$ are given as
\begin{equation}
  \frac{{\partial {h^2}}}{{{\partial ^2}{k_i}}} = \frac{{Aa(a - 1)b\left[ {b - {{\left( {{k_i}} \right)}^a}} \right]{{\left( {{k_i}} \right)}^{a - 2}}}}{{{{\left[ {\left( {1 + c} \right){{\left( {{k_i}} \right)}^a} + b} \right]}^3}}},\forall i \in \left\{ {1,2, \cdots ,N} \right\}
\end{equation}
\begin{equation}
 \frac{{\partial {h^2}}}{{\partial {k_i}\partial {k_j}}} = 0,\forall i,j \in \left\{ {1,2, \cdots ,N} \right\},i \ne j
\end{equation}

We have $a - 1 < 0$, where $a = 0.7624$, and $b - {\left( {{k_i}} \right)^a} > 0$ for ${k_i} \in \left[ {1,830} \right]$, where $b = 168.2$. Due to the limitation specified in (5.c), ${k_i} \in \left[ {1,830} \right]$ is sufficient enough to cover the whole feasible region, thus we don't specified ${k_i} \in \left[ {1,830} \right]$ in the rest of this paper. Then, it is apparent that $\frac{{\partial {h^2}}}{{{\partial ^2}{k_i}}} < 0,\forall i \in \left\{ {1,2, \cdots ,N} \right\}$. As a result the odd-order partial derivation of ${\rm \textbf{H}}$ is always negative, the even-order partial derivation of ${\rm \textbf{H}}$ is always positive, and consequently (12.a) is a concave function. Thus according to Karush-Kuhn-Tucker (KKT) condition, ${\cal K} = \left[ {\bar k, \cdots ,\bar k} \right]$ is the optimal solution of OPT3.

As last, we use the BBM to solve OPT2 on the basis of the optimal solution of OPT3. BBM is a common solution to solve INLP problem and is suitable to solve OPT2 because of the concavity of our objective function in (12.a), and the optimal solutions of a series of continuous relaxation problems bought by BBM can be easily obtained by using LMA. The details of this procure will be presented in Section V.

\renewcommand\arraystretch{2} 
\begin{table*}
  \centering
  \caption{Spectrum Efficiency (SE) of Two WLANs with Overlapped Channels}\label{1}
  \begin{tabular}{ccc}
\shline \textbf{The Number of Overlapped Channels} & \textbf{Possible Channel Allocation Schemes} & \textbf{SE} \\
\hline
\multirow{6}*{$\left| {\cal X} \right| = 1$}  & ${\textbf{\emph{f}}_1}:{f_i} = \left\{ {\tilde 1} \right\},{f_j} = \left\{ {\tilde 1} \right\}$ & $\eta \left( {{\textbf{\emph{f}}_1}} \right){\rm{ = }}\frac{{2\lambda L}}{{20\left[ {1{\rm{ + }}2\rho \left( 1 \right)} \right]}}{\rm{ = }}\frac{{\lambda L}}{{10\left[ {1{\rm{ + }}2\rho \left( 1 \right)} \right]}}$ \\
\cline{2-3}     & ${\textbf{\emph{f}}_2}:{f_i} = \left\{ {\tilde 1} \right\},{f_j} = \left\{ {\tilde 1,2} \right\}$ & $\eta \left( {{\textbf{\emph{f}}_2}} \right){\rm{ = }}\frac{{2\lambda L}}{{40\left[ {1{\rm{ + }}\rho \left( 1 \right){\rm{ + }}\rho \left( 2 \right)} \right]}}{\rm{ = }}\frac{{\lambda L}}{{20\left[ {1{\rm{ + }}\rho \left( 1 \right){\rm{ + }}\rho \left( 2 \right)} \right]}}$ \\
\cline{2-3}     & ${\textbf{\emph{f}}_3}:{f_i} = \left\{ {\tilde 1} \right\},{f_j} = \left\{ {1,\tilde 2} \right\}$ & $\eta \left( {{\textbf{\emph{f}}_3}} \right){\rm{ = }}\frac{{3\lambda L{\rm{ + }}2\rho \left( 1 \right)\lambda L}}{{40\left[ {1{\rm{ + 2}}\rho \left( 1 \right){\rm{ + }}\rho \left( 2 \right){\rm{ + }}{\rho ^2}\left( 1 \right)} \right]}}$ \\
\cline{2-3}     & ${\textbf{\emph{f}}_4}:{f_i} = \left\{ {\tilde 1} \right\},{f_j} = \left\{ {\tilde 1,23,4} \right\}$ & $\eta \left( {{\textbf{\emph{f}}_4}} \right){\rm{ = }}\frac{{2\lambda L}}{{80\left[ {1{\rm{ + }}\rho \left( 1 \right){\rm{ + }}\rho \left( 4 \right)} \right]}}{\rm{ = }}\frac{{\lambda L}}{{40\left[ {1{\rm{ + }}\rho \left( 1 \right){\rm{ + }}\rho \left( 4 \right)} \right]}}$ \\
\cline{2-3}     & ${\textbf{\emph{f}}_5}:{f_i} = \left\{ {\tilde 1} \right\},{f_j} = \left\{ {1,\tilde 2,3,4} \right\}$ & $\eta \left( {{\textbf{\emph{f}}_5}} \right){\rm{ = }}\frac{{3\lambda L + 2\rho \left( 1 \right)\lambda L}}{{80\left[ {1{\rm{ + }}\rho \left( 1 \right){\rm{ + }}\rho \left( 2 \right) + \rho \left( 4 \right) + {\rho ^2}\left( 1 \right)} \right]}}$ \\
\cline{2-3}     & ${\textbf{\emph{f}}_6}:{f_i} = \left\{ {\tilde 1} \right\},{f_j} = \left\{ {1,2,\tilde 3,4} \right\}$ & $\eta \left( {{\textbf{\emph{f}}_6}} \right){\rm{ = }}\frac{{3\lambda L{\rm{ + }}\rho \left( 1 \right)\lambda L{\rm{ + }}\rho \left( 2 \right)\lambda L}}{{80\left[ {1{\rm{ + }}\rho \left( 1 \right){\rm{ + }}\rho \left( 2 \right){\rm{ + }}\rho \left( 4 \right) + \rho \left( 1 \right)\rho \left( 2 \right)} \right]}}$ \\
\hline
\multirow{3}*{$\left| {\cal X} \right| = 2$} & ${\textbf{\emph{f}}_7}:{f_i} = \left\{ {\tilde 1,2} \right\},{f_j} = \left\{ {\tilde 1,2} \right\}$ & $\eta \left( {{\textbf{\emph{f}}_7}} \right){\rm{ = }}\frac{{2\lambda L}}{{40\left[ {1{\rm{ + }}2\rho \left( 2 \right)} \right]}}{\rm{ = }}\frac{{\lambda L}}{{20\left[ {1{\rm{ + }}2\rho \left( 2 \right)} \right]}}$ \\
\cline{2-3}    & ${\textbf{\emph{f}}_8}:{f_i} = \left\{ {\tilde 1,2} \right\},{f_j} = \left\{ {\tilde 1,2,3,4} \right\}$ & $\eta \left( {{\textbf{\emph{f}}_8}} \right){\rm{ = }}\frac{{2\lambda L}}{{80\left[ {1{\rm{ + }}\rho \left( 2 \right){\rm{ + }}\rho \left( 4 \right)} \right]}}{\rm{ = }}\frac{{\lambda L}}{{80\left[ {1{\rm{ + }}\rho \left( 2 \right){\rm{ + }}\rho \left( 4 \right)} \right]}}$ \\
\cline{2-3}    & ${\textbf{\emph{f}}_9}:{f_i} = \left\{ {\tilde 1,2} \right\},{f_j} = \left\{ {1,2,\tilde 3,4} \right\}$ & $\eta \left( {{\textbf{\emph{f}}_9}} \right){\rm{ = }}\frac{{3\lambda L{\rm{ + }}2\rho \left( 2 \right)\lambda L}}{{80\left[ {1{\rm{ + 2}}\rho \left( 2 \right){\rm{ + }}\rho \left( 4 \right){\rm{ + }}{\rho ^2}\left( 2 \right)} \right]}}$ \\
\hline
$\left| {\cal X} \right| = 4$  & ${\textbf{\emph{f}}_{10}}:{f_i} = \left\{ {\tilde 1,2,3,4} \right\},{f_j} = \left\{ {\tilde 1,2,3,4} \right\}$ & $\eta \left( {{\textbf{\emph{f}}_{10}}} \right){\rm{ = }}\frac{{2\lambda L}}{{80\left[ {1{\rm{ + }}2\rho \left( 4 \right)} \right]}}{\rm{ = }}\frac{{\lambda L}}{{40\left[ {1{\rm{ + }}2\rho \left( 4 \right)} \right]}}$ \\
\shline
\end{tabular}

\end{table*}

\emph{2) When the number of WLANs $N$ is larger than the number of available channels $K$ (i.e., $N > K$):}

If the network has more than 4 WLANs and we do not have enough channels to allocate each WLAN non-overlapped channels, there must be overlapped channels between WLANs. An interesting issue is that how these WLANs overlap can get the optimal throughput. In this case, we use the term Spectrum Efficiency (SE) to measure the expected network throughput per frequency spectrum, that is
\begin{equation}
  \eta \left( \textbf{\emph{f}} \right){\rm{ = }}\frac{{T{h_{{{\cal O}_{\cal X}}}}\left( \textbf{\emph{f}} \right)}}{{B{W_{{{\cal O}_{\cal X}}}}\left( \textbf{\emph{f}} \right)}}
\end{equation}
where $T{h_{{{\cal O}_{\cal X}}}}\left( \textbf{\emph{f}} \right)$ and $B{W_{{{\cal O}_{\cal X}}}}\left( \textbf{\emph{f}} \right)$ is the achieved throughput and the total bandwidth been used of an overlapped set, ${{\cal O}_{\cal X}}$, under channel allocation scheme $\textbf{\emph{f}}$ respectively.

When there are overlapped channels among WLANs, we have ${\cal F} = \left\{ {\textbf{\emph{f}}:O\left( \textbf{\emph{f}} \right){\rm{ = 1}}} \right\}$, the ways they can overlap with each other will directly influence the interactions of WLANs. Hence, it’s important to find an efficient way of overlapping to maximize the throughput per spectrum frequency. We want to prove that for a set of WLANs with overlapped channels, ${{\cal O}_{\cal X}}$, the channel allocation scheme with $\left| {\cal X} \right| = 1$ can get the optimal SE for WLANs in this set. The following analysis is based on the mathematical induction.

i) We first consider there are two WLANs in this set, i.e., ${{\cal O}_{\cal X}} = \left\{ {i,j} \right\},\left| {{{\cal O}_{\cal X}}} \right| = 2$. There are three ways that WLAN $i$ and $j$ can overlap with each other based on the number of channels they overlapped, that is $\left| {\cal X} \right|$.

At first, we assume that the number of overlapped channels is one between WLAN $i$ and $j$, and we also assume that WLAN $i$ has been allocated the overlapped channel (i.e., channel (1)). Under these assumptions, there are six schemes that WLAN $i$ and $j$ can overlap with each other, we calculate SE of each possible scheme, as shown in the first line of Table I. Then, we consider the case that the number of overlapped channels is two between WLAN $i$ and $j$ (i.e., channel (1) and (2)). All the SEs are shown in the second line of Table I \footnote{Although the acquisitions of primary channels of WLAN $i$ and $j$ may be different, they can get the same set of feasible network states, we classify them as the same type (i.e., $\textbf{\emph{f}}:{f_i} = \left\{ {\tilde 1,2} \right\},{f_j} = \left\{ {\tilde 1,2,3,4} \right\}$ and $\textbf{\emph{f}}:{f_i} = \left\{ {\tilde 1,2} \right\},{f_j} = \left\{ {1,\tilde 2,3,4} \right\}$.}. Finally, if the number of overlapped channels is four between WLAN $i$ and $j$ (i.e., channel (1) to (4)), there is no way they can transmit together regardless of the position of primary channel, because the selected channel for transmission of each WLAN is the largest contiguous subset of available channels that contains the primary channel under DCB. Thus there is only one scheme corresponding to this situation, as shown in the last line of Table I.

We compare all SEs shown in Table I and find out that the scheme ${\textbf{\emph{f}}_1}:{f_i} = \left\{ {\tilde 1} \right\},{f_j} = \left\{ {\tilde 1} \right\}$ with $\left| {\cal X} \right| = 1$ can get the best SE of WLANs in this set, ${{\cal O}_{\cal X}}$.

ii) We next assume our conclusion stands when there are $M$ WLANs overlap with each other, $\left| {{{\cal O}_{\cal X}}} \right|{\rm{ = }}M$, which means there is a scheme with $\left| {\cal X} \right| = 1$ can get the optimal SE of WLANs in this set, ${{\cal O}_{\cal X}}$.

iii) Finally we should prove that our conclusion works when there are $M{\rm{ + }}1$ WLANs overlap with each other, $\left| {{{\cal O}_{\cal X}}} \right|{\rm{ = }}M + 1$, which means there exists a scheme with $\left| {\cal X} \right| = 1$ can get the optimal SE of WLANs in this set, ${{\cal O}_{\cal X}}$. We label the newly added WLAN as $WLA{N_m}$. We can treat the original $M$ WLANs as a group, which is using the same single basic channel and primary channel for transmission, $\textbf{\emph{f}}:{f_i} = \left\{ {\tilde 1} \right\},\forall i \in M$, according to ii) the optimal SE for this group can be obtained. We use the same manner as i) to characterize the ways of overlapping between group $M$ and $WLA{N_m}$. We treat them as two WLANs with overlapped channels, i.e., ${{\cal O}_{\cal X}} = \left\{ {M,m} \right\},\left| {{{\cal O}_{\cal X}}} \right| = 2$, then after enumerate and compare all possible overlapping allocation schemes we can obtain there is a scheme with $\left| {\cal X} \right| = 1$ has the optimal SE of WLANs in this set, ${{\cal O}_{\cal X}}$.

According to the above analysis, we can arrive at the conclusion that for an overlapped set, ${{\cal O}_{\cal X}}$, there is a channel allocation scheme with $\left| {\cal X} \right| = 1$ can get the optimal SE of WLANs in this set. Therefore, when $N > K$, we could divide $N$ WLANs into $K$ groups and each group is assigned an independent basic channel. Each group of WLANs is a set of WLANs with one overlapped channel, i.e, ${{\cal O}_{\cal X}},\left| {\cal X} \right| = 1$. Thus the optimal SE of WLANs in each set can be obtained under this channel allocation scheme, and the optimal throughput performance will be obtained with some adjustments of the number of WLANs falls into each group. So that Theorem 1 stands when $N > K$.

Based on Theorem 1, when $N > K$, the channel allocation scheme with the minimum $O\left( \textbf{\emph{f}} \right)$ is $O\left( \textbf{\emph{f}} \right) = 1$, and we divide $N$ WLANs into $K$ groups and let the WLANs in a group use a single basic channel. That is, when $N > K$, the CB technique is forbidden to prevent too much interference among WLANs. Let ${n_k},k \in \left\{ {1, \cdots ,K} \right\}$ denote the number of WLANs that is allocated channel $k$. Each WLAN uses a single basic channel for transmission and the activity ratio of each WLAN is then $\rho \left( 1 \right) = \frac{{E\left[ {T\left( 1 \right)} \right]}}{{E\left[ B \right]}}$. Thus, the throughput achieved by WLANs in a group can be computed as $\frac{{{n_k}\lambda L}}{{1 + \rho \left( 1 \right){n_k}}}$, then OPT1 can be rewritten as the following INLP problem:
\begin{subequations}
\begin{equation}
  {\rm{OPT4:     }}\max {\rm{ }}g\left( {{n_1}, \cdots ,{n_K}} \right) = \sum\limits_{k = 1}^K {\frac{{A{n_k}}}{{1 + B{n_k}}}}
\end{equation}
\begin{equation}
s.t.{\rm{  }}\sum\limits_{k = 1}^K {{n_k} = N}
\end{equation}
\begin{equation}
\qquad \qquad {n_k} \ge 1,{n_k} \in {N^{\rm{ + }}}
\end{equation}
\end{subequations}
where $A{\rm{ = }}\lambda L$ and $B{\rm{ = }}\rho \left( 1 \right)$, which are both constants in our model. The objective function $g\left( {{n_1}, \cdots ,{n_K}} \right)$ in (18a) represents the aggregate network throughput under the grouping scheme ${\cal N} = \left[ {{n_1}, \cdots ,{n_K}} \right]$. Eq.(18b) means that the sum of ${n_k}$ must equal to the total number of WLANs $N$, and Eq.(18c) means ${n_k}$ must be a positive integer. After solving OPT4 we can get an optimal grouping scheme, ${{\cal N}^ * } = \left[ {n_1^ * , \cdots ,n_K^ * } \right]$, and the optimal channel allocation scheme can be obtained by setting ${f_i} = \left\{ {\tilde k} \right\}$ if $i \in {k^{th}} \ {\rm{ }}group$ and letting channel $k$ be the primary channel of $WLA{N_i}$.

In order to solve OPT4, we first relax ${n_k}$ to a continuous variable. Since the constraint in (18b) is an equation, we can use LMA to solve the relaxation problem of OPT4. Introducing a Lagrange multiplier $\xi $ ($\xi  \ge 0$) to establish a Lagrange function $G\left( {{n_1}, \cdots ,{n_K},\xi } \right)$, which is
\begin{equation}
  G\left( {{n_1}, \cdots ,{n_K},\xi } \right) = \sum\limits_{k = 1}^K {\frac{{B{n_k}}}{{1 + A{n_k}}}}  + \xi \left( {{n_1} +  \cdots  + {n_K} - N} \right)
\end{equation}
then we take the derivatives of $G\left( {{n_1},{n_2},{n_3},{n_4},\xi } \right)$ for each unknown variables, and make them equal to zero to get the extreme point, which can be represented as
\begin{equation}
  \begin{array}{l}
\frac{{\partial G}}{{\partial {n_1}}} = \frac{B}{{{{\left( {1 + A{n_1}} \right)}^2}}} + \xi  = 0\\
{\rm{               }} \vdots \\
\frac{{\partial G}}{{\partial {n_K}}} = \frac{B}{{{{\left( {1 + A{n_K}} \right)}^2}}} + \xi  = 0\\
\frac{{\partial G}}{{\partial \xi }} = {n_1} +  \cdots  + {n_K} - N = 0
\end{array}
\end{equation}

By solving (20) we can get a group scheme ${\cal N}{\rm{ = }}\left\{ {\bar n, \cdots ,\bar n} \right\}$, where $\bar n = N/K$, $\bar n$ can be interpreted as the mean number of WLANs that be allocated the same single basic channel, which might be a non-integer.

The Hessian matrix of (18.a) is
\[{\rm \textbf{H}} = \left\{ {\begin{array}{*{20}{c}}
{\frac{{\partial {g^2}}}{{{\partial ^2}{n_1}}}}&{\frac{{\partial {g^2}}}{{\partial {n_1}\partial {n_2}}}}& \cdots &{\frac{{\partial {g^2}}}{{\partial {n_1}\partial {n_K}}}}\\
{\frac{{\partial {g^2}}}{{\partial {n_2}\partial {n_1}}}}&{\frac{{\partial {g^2}}}{{{\partial ^2}{n_2}}}}& \cdots &{\frac{{\partial {g^2}}}{{\partial {n_2}\partial {n_K}}}}\\
 \vdots & \vdots & \ddots & \vdots \\
{\frac{{\partial {g^2}}}{{\partial {n_K}\partial {n_1}}}}&{\frac{{\partial {g^2}}}{{\partial {n_K}\partial {n_2}}}}& \cdots &{\frac{{\partial {g^2}}}{{{\partial ^2}{n_K}}}}
\end{array}} \right\}\]
the elements of ${\rm \textbf{H}}$ are given as
\begin{equation}
  \frac{{\partial {g^2}}}{{{\partial ^2}{n_k}}} = \frac{{ - 2AB}}{{{{\left( {1 + B{n_k}} \right)}^3}}},\forall k \in \left\{ {1, \cdots ,K} \right\}
\end{equation}
\begin{equation}
  \frac{{\partial {g^2}}}{{\partial {n_i}\partial {n_j}}} = 0,\forall i,j \in \left\{ {1, \cdots ,K} \right\},i \ne j
\end{equation}

It’s obviously $\frac{{\partial {g^2}}}{{{\partial ^2}{n_k}}} < 0$, so (18.a) is a concave function, and the extreme point, ${\cal N}{\rm{ = }}\left\{ {\bar n, \cdots ,\bar n} \right\}$ is also the maximum point. Finally, we can use the BBM to get the optimal solution of OPT4. The details of this procure will be presented in Section V.

\renewcommand\arraystretch{1.5}
\begin{table}
  \centering
  \caption{An Example of the Channel Allocation Algorithm}\label{2}
  \begin{tabular}{ccccc}
\shline \quad & $\left\{ {{k_1},{k_2},{k_3}} \right\}$ & \textbf{Feasibility} & \textbf{System Throughput} & \makecell{\textbf{[Lower bound,} \\ \textbf{Upper bound}]}\\
\hline ${1^{st}}$  & \makecell{$\left\{ {1,1,1} \right\}$ \\ $\left\{ {\frac{7}{3},\frac{7}{3},\frac{7}{3}} \right\}$} & \makecell{Yes \\ No} & \makecell{186.8310 \\ 358.8981} & \makecell{[186.8310, \\ 358.8981]}\\
\hline ${2^{nd}}$  & \makecell{$\left\{ {2,\frac{5}{2},\frac{5}{2}} \right\}$ \\ $\left\{ {4,\frac{3}{2},\frac{3}{2}} \right\}$} & \makecell{NO \\ No} & \makecell{358.5351 \\ 350.7984} & \makecell{[186.8310, \\ 378.2528]}\\
\hline ${3^{rd}}$  & \makecell{$\left\{ {2,2,3} \right\}$ \\ $\left\{ {2,4,1} \right\}$} & \makecell{No \\ Yes} & \makecell{357.5556 \\ 339.8579} & \makecell{[339.8579, \\ 357.5556]}\\
\hline ${4^{th}}$  & \makecell{$\left\{ {2,2,2} \right\}$ \\ $\left\{ {2,2,4} \right\}$} & \makecell{Yes \\ No} & \makecell{343.7781 \\ /} & \makecell{[343.7781, \\ 357.5556]}\\
\hline \textbf{Result}  & \makecell{$\left\{ {2,2,2} \right\}$} & \makecell{Yes} & \makecell{343.7781} & / \\
\shline
\end{tabular}

\end{table}

\section{Optimal Channel Allocation Algorithm Design}
Based on the performance analysis and the constructed INIP models in Section IV, we propose a channel allocation algorithm based on the BBM to solve OPT2 and OPT4 to get the optimal channel allocation scheme, as well as the maximal aggregate throughput of the DCB network.
\subsection{Proposed Channel Allocation Algorithm}
Algorithm 1 presents the pseudo-code to find the optimal channel allocation scheme ${\textbf{\emph{f}}^ * }$ when $N \le K$. When $N > K$, we only need to change the adjustable variable in Algorithm 1 to ${n_k},k \in \left\{ {1, \cdots ,K} \right\}$ and set the initial state as $\bar n = N/K$, $L = g\left( {1, \cdots ,1} \right)$, $U = g\left\{ {\bar n, \cdots ,\bar n} \right\}$, where $L$ and $U$ are the lower bound and upper bound respectively. Then similar procedure can be performed to solve OPT4. It is important to note that the proposed channel allocation algorithm only requires the information of $N$ and $K$ while no information exchange is required among WLANs.
\renewcommand\arraystretch{1.5}
\begin{table}[t]
  \centering
  \caption{An Example of the Greedy Scheme}\label{3}
  \begin{tabular}{cccc}
\shline \quad & $\left\{ {{k_1},{k_2},{k_3}} \right\}$ & \textbf{Feasibility} & \textbf{System Throughput} \\
\hline ${1^{st}}$  & \makecell{$\left\{ {1,1,1} \right\}$} & \makecell{Yes} & \makecell{186.8310}  \\
\hline ${2^{nd}}$  & \makecell{$\left\{ {2,1,1} \right\}$} & \makecell{Yes} & \makecell{239.1467} \\
\hline ${3^{rd}}$  & \makecell{$\left\{ {4,1,1} \right\}$} & \makecell{Yes} & \makecell{287.5422} \\
\hline ${4^{th}}$  & \makecell{$\left\{ {8,1,1} \right\}$} & \makecell{No} & \makecell{/}\\
\hline ${5^{th}}$  & \makecell{$\left\{ {4,2,1} \right\}$} & \makecell{Yes} & \makecell{339.8579}\\
\hline \textbf{Result}  & \makecell{$\left\{ {4,2,1} \right\}$} & \makecell{Yes} & \makecell{339.8579}  \\
\shline
\end{tabular}

\end{table}

In Algorithm 1, steps 4-5 show the optimal solution obtained by solving relaxation problems through LMA. In steps 7-22, between two branches, we set the feasible objective value of the feasible branch as the new lower bound, and if there is a higher objective value of an infeasible solution, we set it as the new upper bound, then keep branching under this solution until we find an optimal feasible solution that maximizes the lower bound. Algorithm 1 can be summarized as following:

Step 1: Initialization.

Steps 2-6: For a variable, get two branches by adding the constraints ${k_i} \le \left\lfloor {\bar k} \right\rfloor  = {2^m}$ and ${k_i} \ge \left\lfloor {\bar k} \right\rfloor  + {2^{m + 1}}$ to OPT3 respectively ($\left\lfloor {\bar k} \right\rfloor $ is the round down integer value of ${k_i}$, and it is a multiple of 2), or by adding the constraints ${n_k} \le \left\lfloor {\bar n} \right\rfloor $ and ${n_k} \ge \left\lfloor {\bar n} \right\rfloor  + 1$ to OPT4 respectively ($\bar n$ is the round down integer value of ${n_k}$).

Steps 7-27: Examine the feasibility of each branch to see whether it is a feasible integer solution (satisfy the constraints in OPT2 or in OPT4). Update $L$ and $U$ accordingly.

Step 28: Output the optimal channel allocation scheme, ${\textbf{\emph{f}}^ * }$, and the maximal aggregate throughput, $T{h^ * }$.

\renewcommand\arraystretch{1.5}
\begin{table*}[t]
  \centering
  \caption{Comparisons Between the Proposed Algorithm and the Greedy Scheme}\label{4}
  \begin{tabular}{ccccccc}
\shline \multicolumn{2} {c} {\textbf{Parameters}} & \multicolumn{4} {c} {\textbf{Aggregate Throughput(Mbps)}} & \textbf{JFI}\\
\hline Allocating scheme  & $\left\{ {{k_A},{k_B},{k_C}} \right\}$ & WLAN A & WLAN B & WLAN C & SUM &   \\
\hline  The Proposed Algorithm  & $\left\{ {2,2,2} \right\}$ & 114.5927 & 114.5927 & 114.5927 & 343.7780 & 1 \\
\hline The Greedy Scheme  & $\left\{ {4,2,1} \right\}$ & 162.9881 & 114.5927 & 62.2770 & 339.9578 & 0.8836 \\
\hline \textbf{Gain}  & / & 0.4223 & 0 & 0.4565 & / & / \\
\shline
\end{tabular}

\end{table*}

To better illustrate the proposed algorithm, we give the procedure under a network setting with $N = 3,K = 7$. In Table II the initial lower bound is set to $h\left( {1,1,1} \right)$, the initial upper bound is set to $h'\left( {7/3,7/3,7/3} \right)$, which is the optimal solution for OPT3. We select the variable to branch in an ascending order (i.e., ${h_1},{h_2},{h_3}$). Then we update the lower bound and upper bound according to Algorithm 1 and eventually find out the optimal channel allocation scheme. Note that if the sum of assigned channels in the grouping scheme exceeds $K$, we treat this grouping scheme as an invalid scheme and its objective value is set to zero.

Due to the convexity of the objective function of OPT3 and OPT4, the channel allocation algorithm based on the BBM yields a solution that in general is the global optimal solution \cite{gupta1985branch}. The computations of the algorithm is simple and the complexity is $O\left( N \right)$ if $N \le K$ or $O\left( K \right)$ if $N > K$. Thus, the proposed channel allocation algorithm can converge to the optimal solution quickly.

\begin{table}
  \centering
  \caption{Parameters Values Based on Ieee 802.11ac}\label{5}
  \begin{tabular}{ccc}
\shline \textbf{Parameter} & \textbf{Notation} & \textbf{Value}\\
\hline Packet length & ${L_d}$ & 12000bits\\
       Number of aggregated packets & ${K_A}$ & 64 packets\\
       Contention window & $CW$ & 16 slots\\
       Slot duration & ${T_{slot}}$ & 9$\mu s$\\
       Average backoff duration &  $E\left( B \right)$ &$CW\frac{{{T_{slot}}}}{2}$\\
\shline
\end{tabular}

\end{table}

\begin{table}
  \centering
  \caption{Transmission Duration for Different Channel Number in Use}\label{6}
  \begin{tabular}{ccccc}
  \shline
      $k'$ & $\varepsilon \left( {k'} \right)$ & $M$ & $R$ & $T\left( {k'} \right)$ \\
  \hline
      1 & 52 & 6 bits(64-QAM) & 5/6 & 12.26ms \\
      2 & 108 & 6 bits(64-QAM) & 3/4 & 6.63ms \\
      3 & 234 & 4 bits(16-QAM) & 3/4 & 4.64ms \\
      4 & 468 & 4 bits(16-QAM) & 1/2 & 3.52ms \\
  \shline
  \end{tabular}
\end{table}

Table III shows an example of the greedy scheme, each variable attempts to maximize their own interest at each iteration. We set the initial state as ${\cal K}{\rm{ = }}\left\{ {1,1,1} \right\}$, when $N \le K$, we choose a variable in an ascending order and keeping doubling the number of channels assigned to it while make sure the number of available basic channels does not exceed $K$ (i.e., ${\cal K}{\rm{ = }}\left\{ {4,2,1} \right\}$ when $N = 3,K = 7$). The final objective values are shown in the last row of two Table II and III, the proposed algorithm shows better system throughput than the greedy scheme. An interesting question comes into mind, in the greedy scheme, the whole number of available basic channels come in handy, why achieves a lower system throughput? We will explain this phenomenon in the next part.

The procedures to find a grouping scheme of the proposed algorithm and the greedy scheme when $N > K$ are similar to the above. Only in the greedy scheme, we increase the number of WLANs in the group by one value until the number of WLANs allowed is reached. For example, in the network settings with $N = 7,K = 3$, the grouping scheme obtained by the proposed algorithm is ${\cal N} = \left\{ {2,2,3} \right\}$ and the one obtained by the greedy scheme is ${\cal N} = \left\{ {5,1,1} \right\}$, we have $g\left( {2,2,3} \right) > g\left( {5,1,1} \right)$.

\subsection{Case of Interest}
We consider a scene that there are 3 neighboring WLANs and 7 basic channels available. Our grouping scheme obtained by the proposed algorithm is ${\cal K}{\rm{ = }}\left\{ {2,2,2} \right\}$ and the corresponding channel allocation scheme is $\textbf{\emph{f}}:{f_A} = \left\{ {\tilde 1,2} \right\},{f_B} = \left\{ {\tilde 3,4} \right\},{f_C} = \left\{ {\tilde 5,6} \right\}$, under this scheme, there is only 6 channels used for transmission. However, the grouping scheme obtained by the greedy scheme is ${\cal K}{\rm{ = }}\left\{ {4,2,1} \right\}$ and the corresponding channel allocation scheme is $\textbf{\emph{f}}:{f_A} = \left\{ {\tilde 1,2,3,4} \right\},{f_B} = \left\{ {\tilde 5,6} \right\},{f_C} = \left\{ {\tilde 7} \right\}$, there are 7 channels used for transmission.

We use the simulation parameters present in section VI to compare two solutions in terms of achieved throughput in each WLAN, the network-wide throughput and Jain’s Fairness Index (JFI). What’s more, we use the term \textbf{Gain} to represent the rate of throughput increase of each WLAN between two schemes. For example, the number of channels allocated to $WLA{N_A}$ is 2 in our scheme, and is 4 in the greedy scheme, the Gain of $WLA{N_A}$ is $\left[ {T{h_A}\left( 4 \right) - T{h_A}\left( 2 \right)} \right]/T{h_A}\left( 2 \right)$, in the same way, the Gain of $WLA{N_C}$ is $\left[ {T{h_C}\left( 2 \right) - T{h_C}\left( 1 \right)} \right]/Th\left( 1 \right)$, the number in brackets represent the number of allocated non-overlapped channels ${k_i}$  for $WLA{N_i}$.

From Table IV, we can see although the scheme obtained by the proposed algorithm do not use the total number of available basic channels, but it gets a better network-wide throughput and JFI than the greedy scheme. Through the Gain of each WLAN we can get the reason of this phenomenon, that is the Gain obtained by WLAN A is less than the Gain obtained by WLAN C. From the insight of engineering practice, the duration of some headers and preambles is not affected by the channel width in 802.11ac network. Therefore, doubling the number of a WLAN’s allocated basic channels, the transmission rate of this WLAN can’t boost up to twice that. Indeed, the more number of basic channels is, the less Gain can get by doubling it. What’s more, from the point of view of JFI, allocating each WLAN approximately equal number of non-overlapped basic channels can guarantee a good fairness among WLANs.

\section{Simulation Results}
This section evaluates the performance of our proposed algorithm through simulations. The wireless networking environment is configured as a bulk of WLANs that are all within the carrier-sensing range of each other. Simulation parameters are presented in Table V, given by amendment 802.11ac to keep the error probability ${p_e}$ below 10\%. Using these parameters we can calculate the packet transmission duration $T\left( {k'} \right)$, for each channel number in use $k'$, as shown in Table VI, $M$, $R$ is the modulation and the coding rate respectively, $\varepsilon \left( {k'} \right)$ is the number of data subcarriers when $k'$ basic channels are used.

\subsection{Accuracy Validation of the CTMC Model}
In this part, we first examine the accuracy of the CTMC model. Fig. 7 presents the throughput performance of the illustrating example of Fig. 3 with respect to the Backoff Contention Window ($CW$) duration. The relationship between $\lambda $ and $CW$ is given by $\lambda  = \frac{2}{{CW{T_{slot}}}}$.  Each point of the simulation is the mean value averaged over 1000 simulations. From Fig. 7, we can see the analytical and simulated results match well. More validation of the CTMC model can be found in [9].

\begin{figure}
\begin{minipage}{0.45\linewidth}
  \centerline{\includegraphics[width=1.8in]{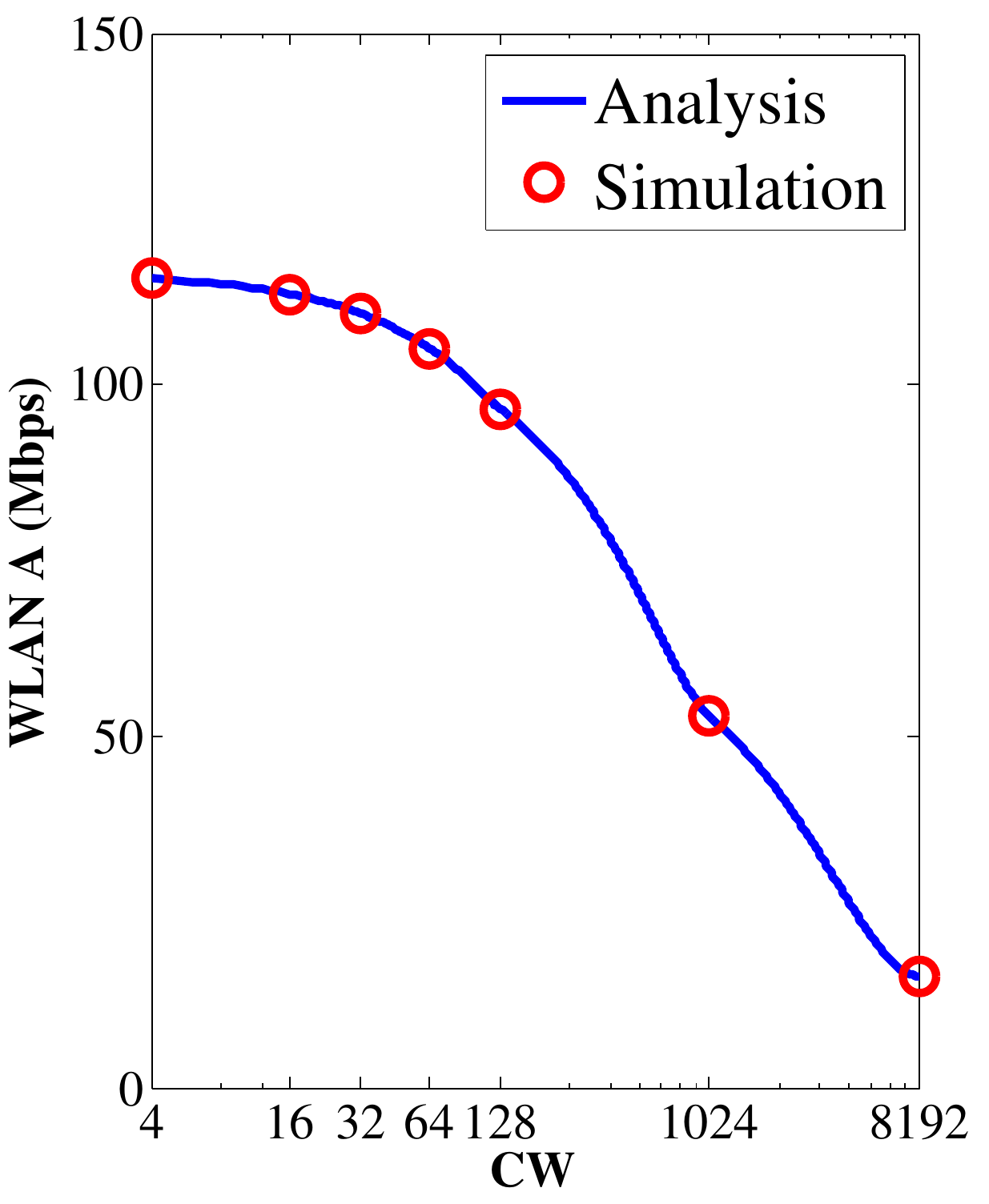}}
  \centerline{(a) WLAN A}
\end{minipage}
\hfill
\begin{minipage}{0.45\linewidth}
  \centerline{\includegraphics[width=1.8in]{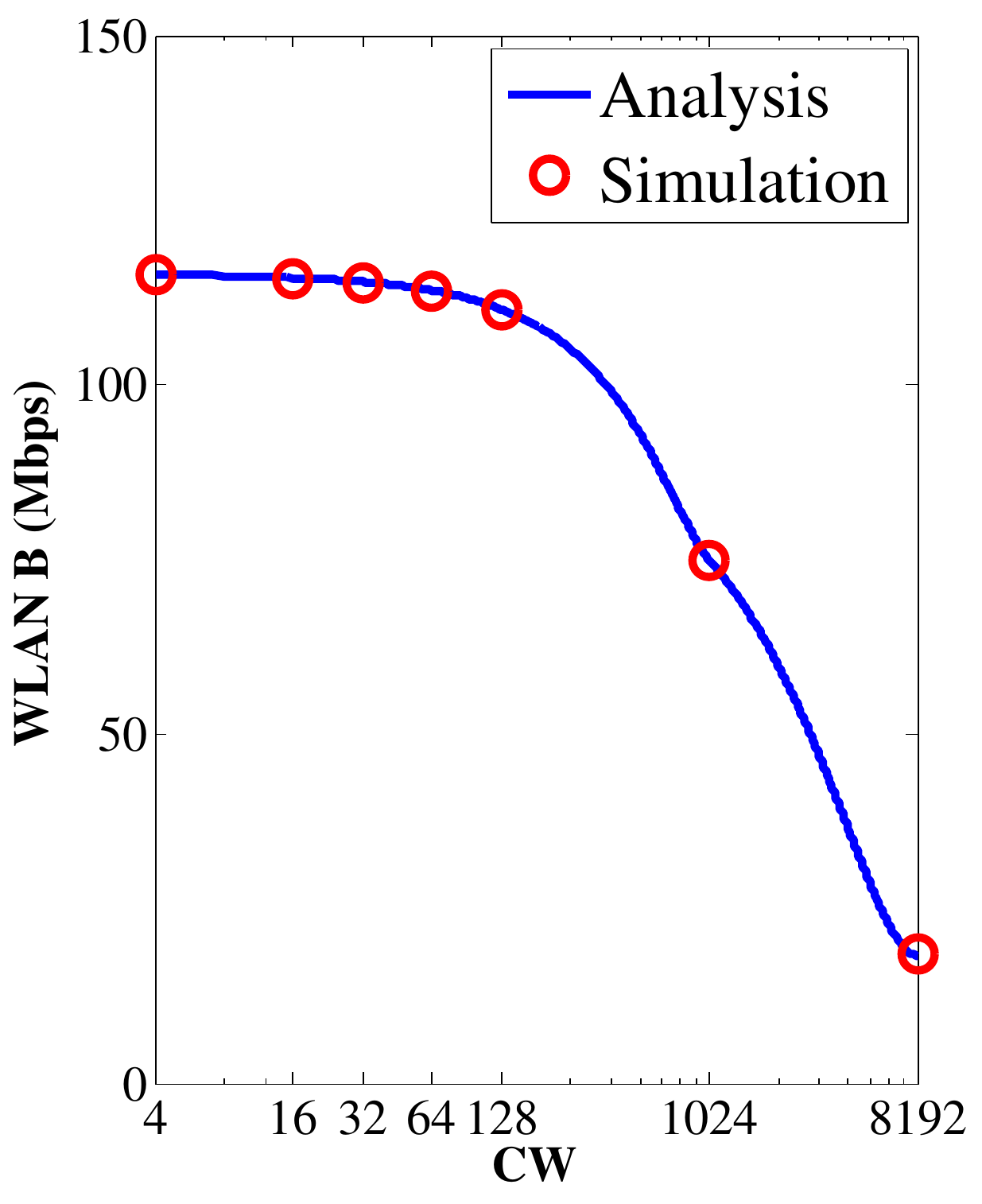}}
  \centerline{(b) WLAN B}
\end{minipage}
\caption{Throughput Comparison between Analysis and Simulation}
\label{7}
\end{figure}

\subsection{Performance of the Proposed Channel Allocation Algorithm}
We next examine the throughput performance when $K = 4$ while $N$ increases from 1 to 10. When $N$ increases from 1 to 4, we use the proposed algorithm to solve OPT2, and when $N$ increases from 5 to 10, we use the proposed algorithm to solve OPT4 to get an optimal channel allocation scheme. We also present the results of a greedy scheme and two random-selection schemes for comparison purpose. In the first random-selection scheme we select a random channel combination for a fixed bandwidth (see Fig. 8(a)), in the other random-selection scheme we select a channel combination with a channel width randomly select from 20MHz to $B{W_{\max }}$ (see Fig. 8(b)). Note that the random-selection of $BW$ increases the opportunities that different WLANs interact with each other because more combinations of the basic channels are possible. Similarly, the aggregate throughput of the random-selection scheme is the mean averaged over 1000 simulations. From Fig. 8, we can see that the proposed algorithm can always converge to the optimal solution, which is obtained by exhaustive search in the feasible region. Moreover, the aggregate throughput achieved by the proposed algorithm is always higher than that of random-selection schemes. It is worth while to note that when $N$ increases from 1 to 4, the throughput achieved by the greedy scheme is the same as that of our proposed algorithm. However, there is a slight drop compared to our proposed algorithm when $N$ increases from 5 to 10. This is because in the objective function of OPT4, $B{\rm{ = }}\rho \left( 1 \right) \gg 1$, we have $\frac{{A{n_k}}}{{1 + B{n_k}}} \approx \frac{A}{B}$. When $N > K$, the optimization of the grouping scheme can only obtain a slight throughput improvement. By contrast, in greedy scheme, when $N > K$, every group attempts to adopt more WLANs to boost its aggregate throughput, while the grouping scheme obtained by the proposed algorithm tends to uniformly allocate the WLANs into different group. Thus, our proposed algorithm can achieve better fairness among WLANs as well as higher aggregate throughputs. The details will be presented later in Fig.10.

\begin{figure}
\begin{minipage}[t]{0.45\linewidth}
  \centerline{\includegraphics[width=1.8in]{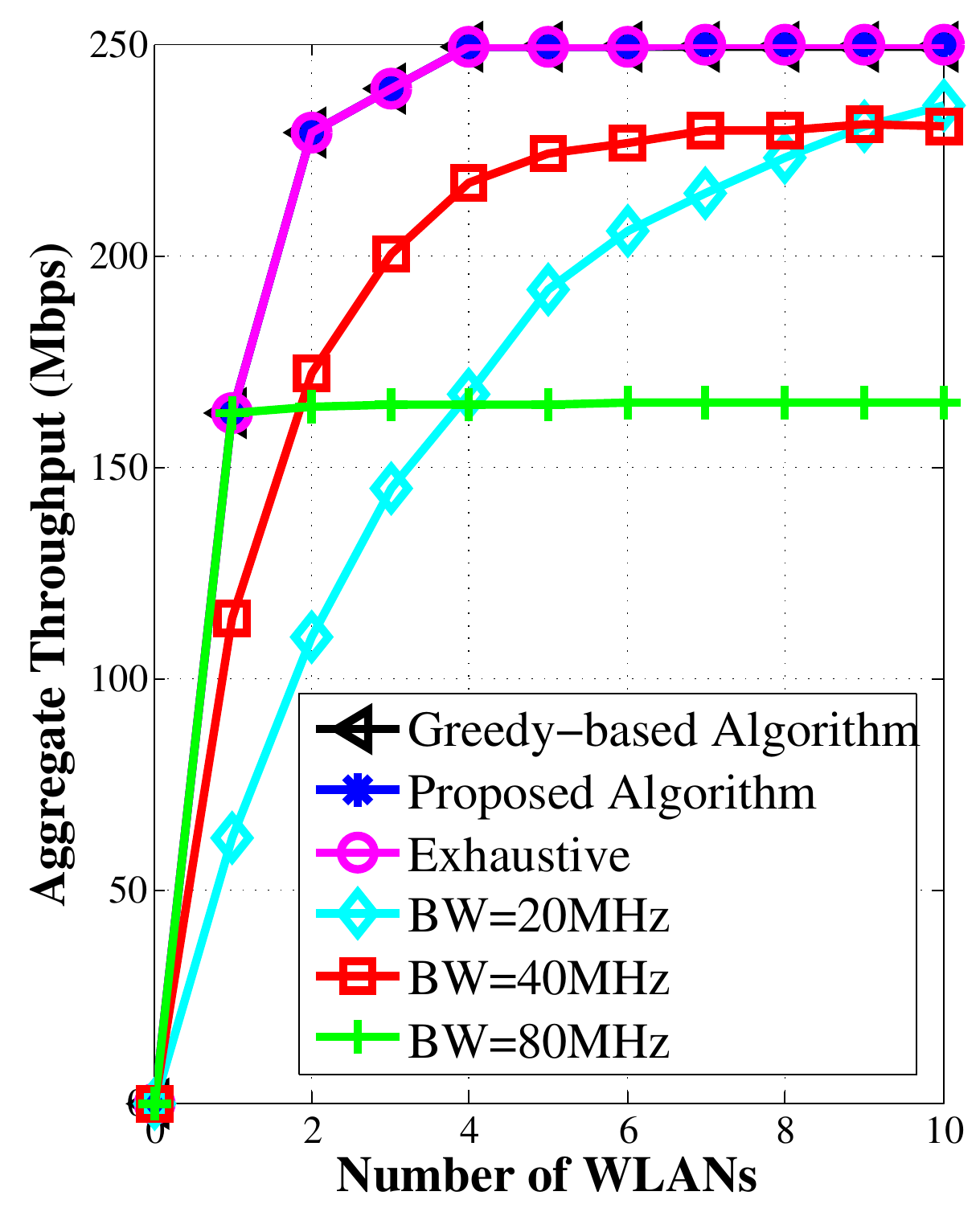}}

  \centerline{(a) Bandwidth Fixed}
  \end{minipage}
  \hfill
  \begin{minipage}[t]{0.45\linewidth}
  \centerline{\includegraphics[width=1.8in]{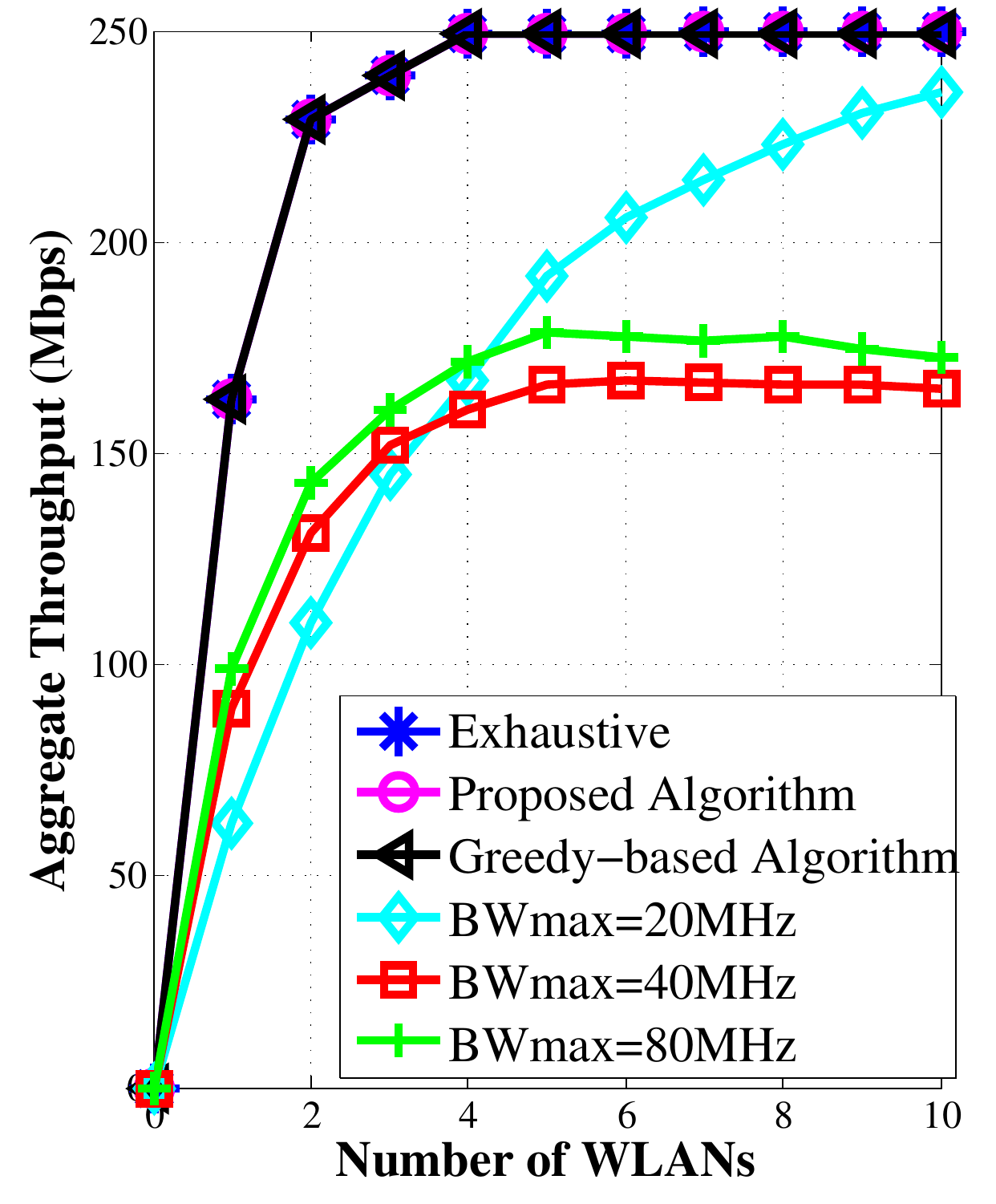}}
  \centerline{(b) Bandwidth Random}
  \end{minipage}
  \caption{Comparison of Aggregate Throughputs when $K = 4$}
  \label{8}
\end{figure}
We next consider a more practical situation of the IEEE 802.11ac WLAN and there are more than 4 available basic channels in 5GHz band, as we can see from Fig. 1. Hence, we set $K = 17$ and increase $N$ from 1 to 20. We present the system performance in terms of aggregate throughput, JFI and Channel Utilization (CU). The JFI is defined as
\begin{equation}
J = \frac{{{{\left( {\sum\nolimits_{i = 1}^N {T{h_i}} } \right)}^2}}}{{N\sum\nolimits_{i = 1}^N {Th_i^2} }}
\end{equation}
\begin{figure*}
\begin{minipage}[t]{0.3\linewidth}
  \centerline{\includegraphics[width=1.8in]{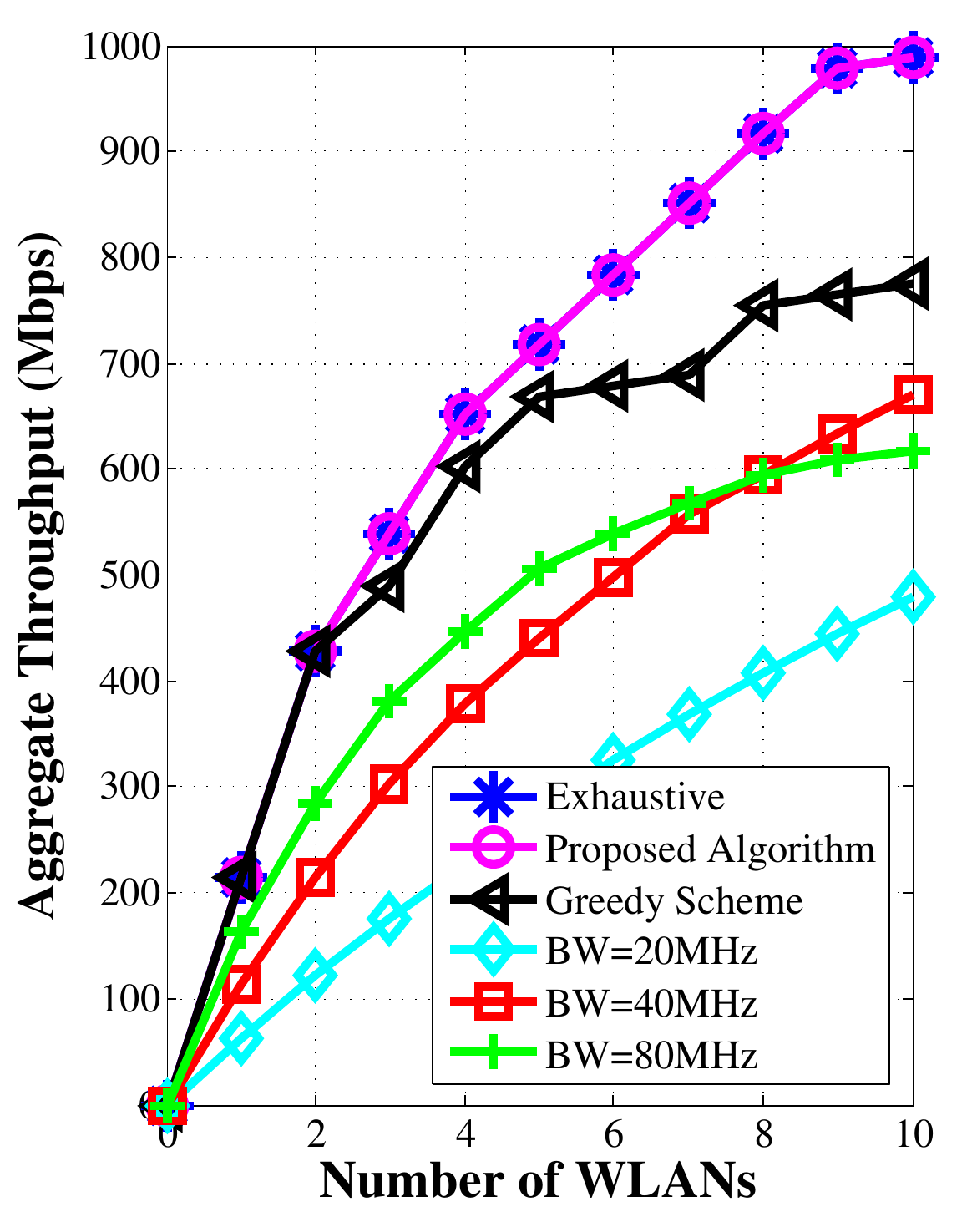}}
  \centerline{(a) Throughput}
\end{minipage}
\hfill
\begin{minipage}[t]{0.3\linewidth}
  \centerline{\includegraphics[width=1.8in]{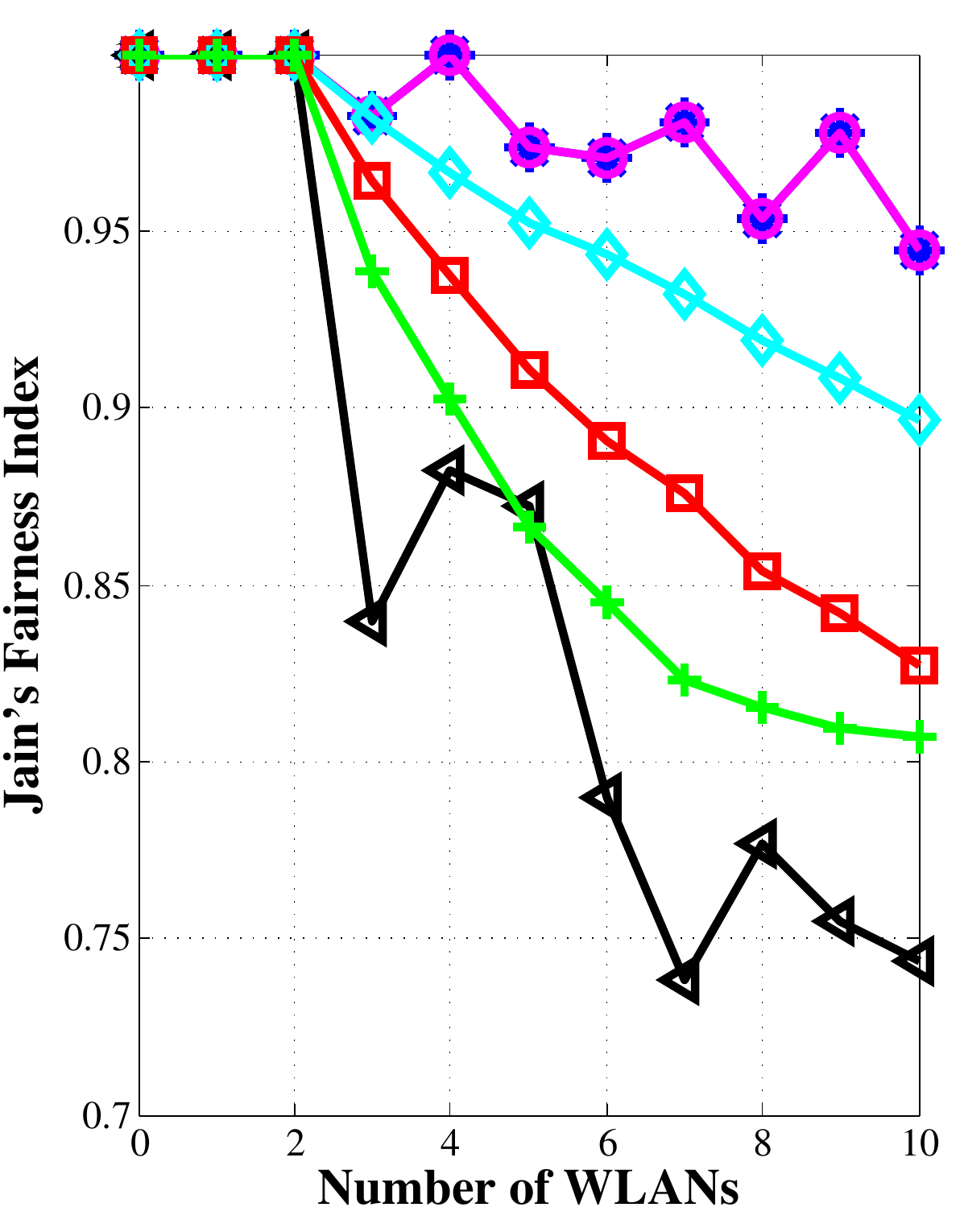}}
  \centerline{(b) Channel Utilization}
\end{minipage}
\hfill
\begin{minipage}[t]{0.3\linewidth}
  \centerline{\includegraphics[width=1.8in]{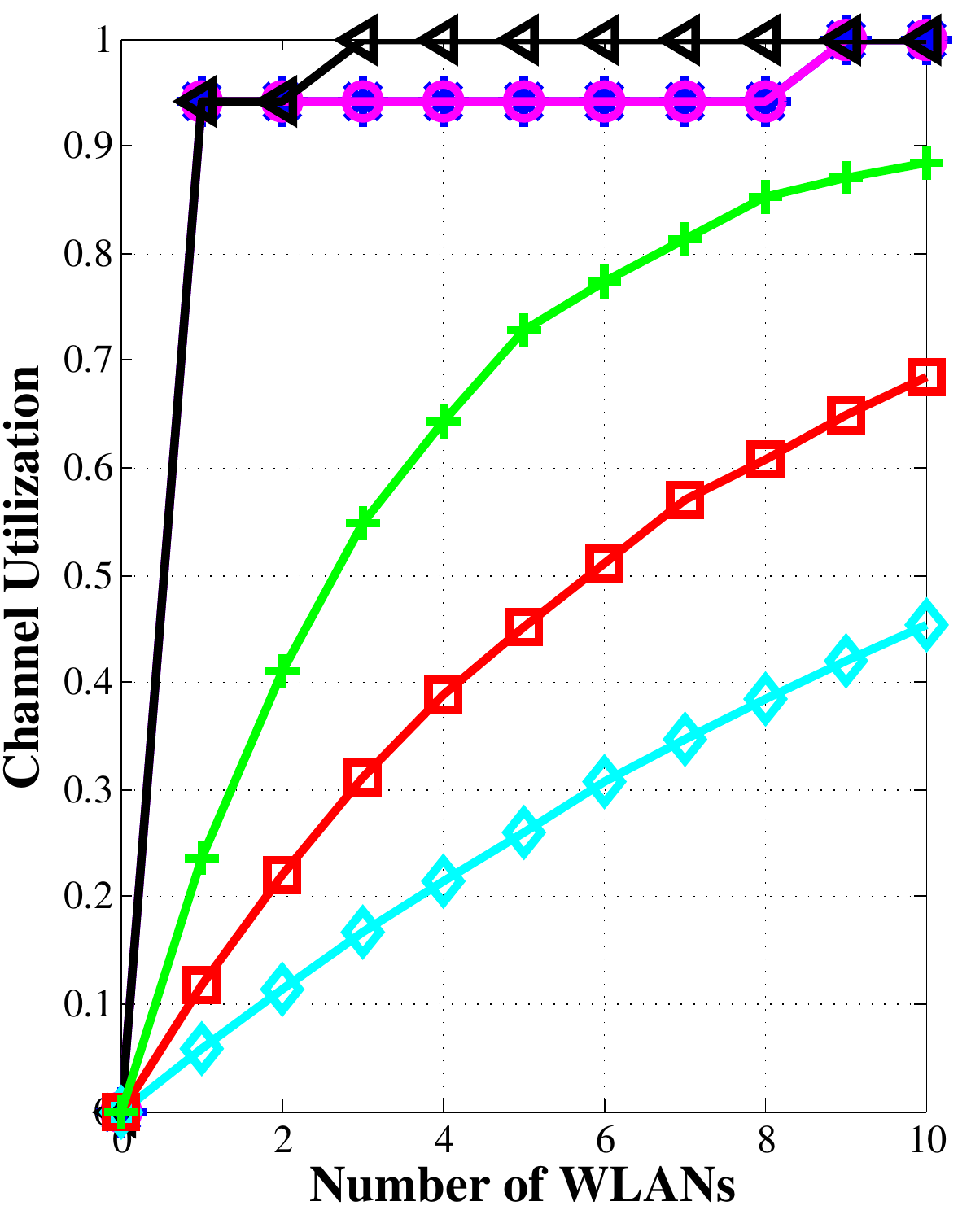}}
  \centerline{(c) Jain's Fairness Index}
\end{minipage}
\caption{Performance Comparison of Different Algorithms with respect to $N$ ($BW$ fixed)}
\label{9}
\end{figure*}

\begin{figure*}
\begin{minipage}[t]{0.3\linewidth}
  \centerline{\includegraphics[width=1.8in]{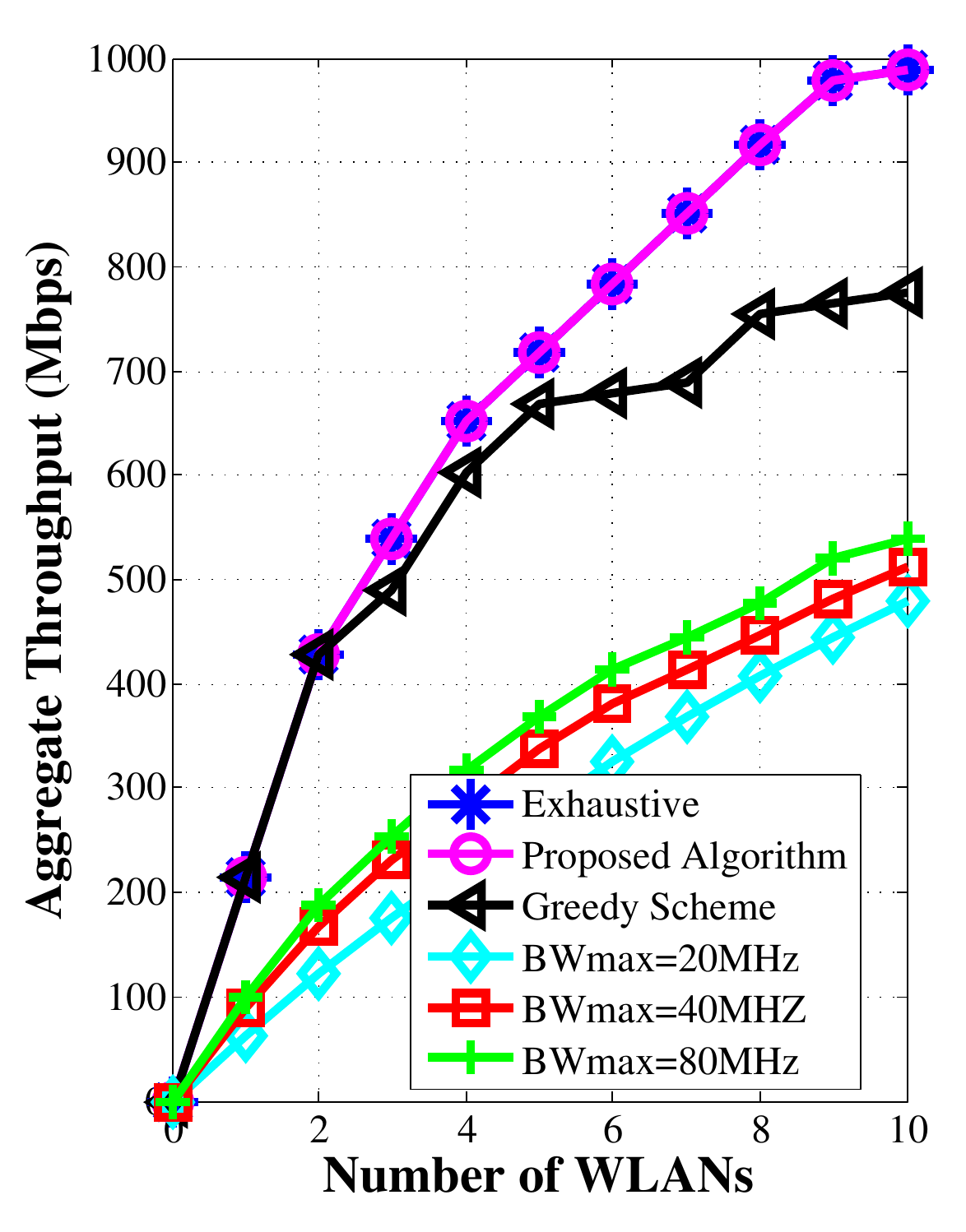}}
  \centerline{(a) Throughput}
\end{minipage}
\hfill
\begin{minipage}[t]{0.3\linewidth}
  \centerline{\includegraphics[width=1.8in]{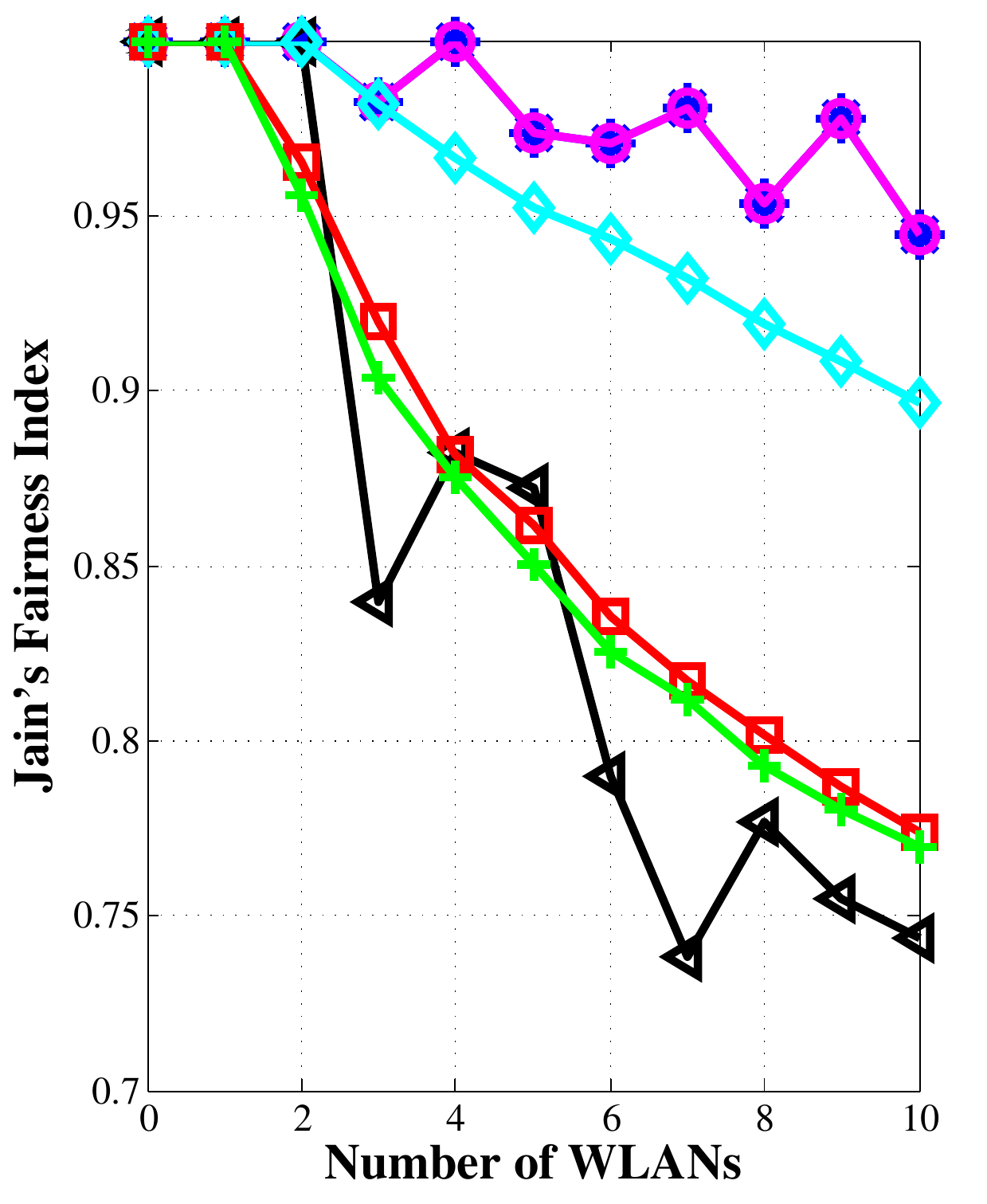}}
  \centerline{(b) Channel Utilization}
\end{minipage}
\hfill
\begin{minipage}[t]{0.3\linewidth}
  \centerline{\includegraphics[width=1.8in]{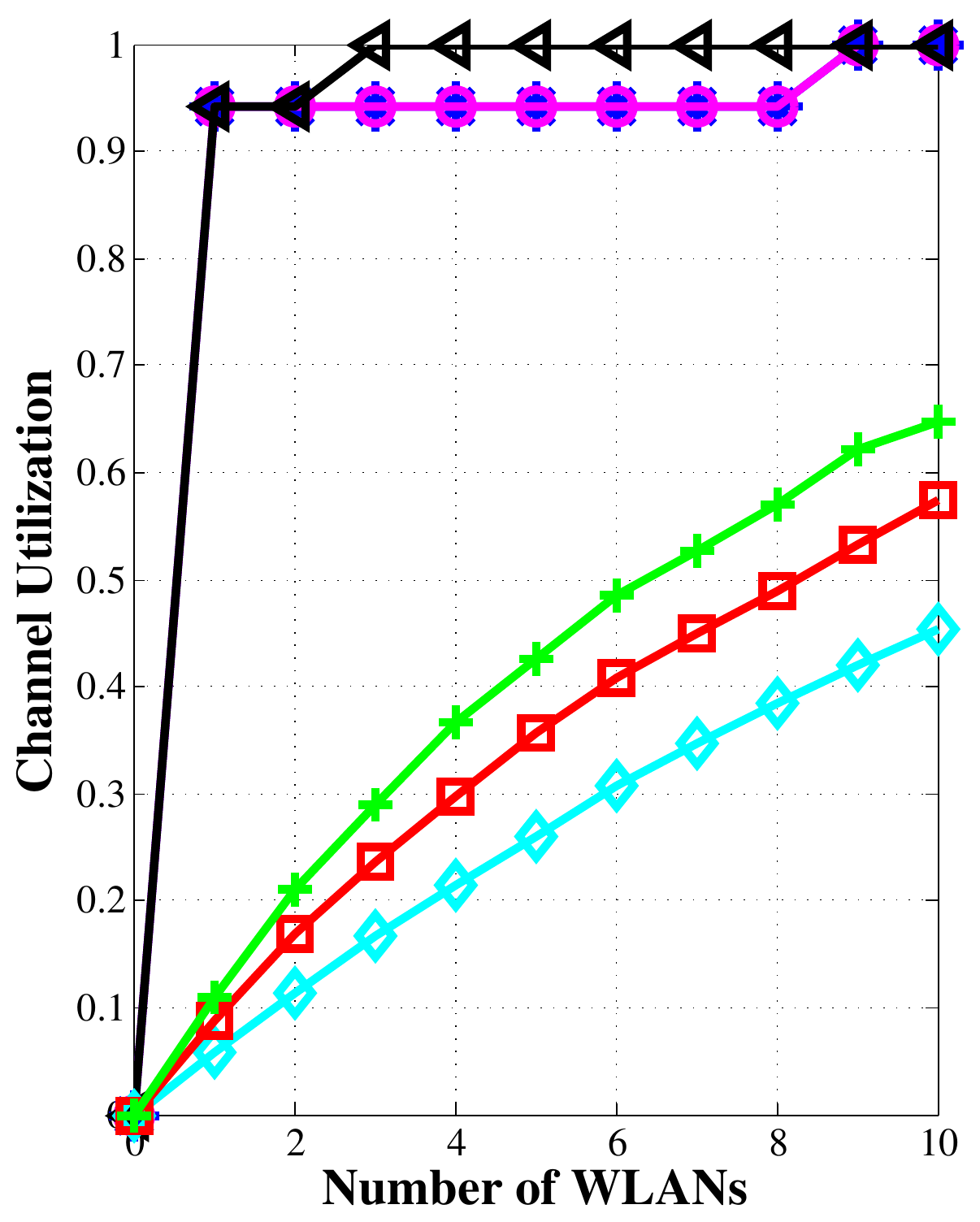}}
  \centerline{(c) Jain's Fairness Index}
\end{minipage}
\caption{Performance Comparison of Different Algorithms with respect to $N$ ($BW$ random)}
\label{10}
\end{figure*}

Moreover, CU is computed as the fraction of basic channels that are occupied by one or more WLANs divided by the total number of basic channels, i.e.,
\begin{equation}
\Gamma \left( f \right) = \frac{1}{K}\sum\limits_{k = 1}^K {{\rm I}\left( k \right)}
\end{equation}
where the ${\rm I}\left( k \right)$ equals to 1 if the basic channel $k$ is found occupied at least one WLAN.

When $N$ increases from 1 to 10, we compare the proposed algorithm with the optimal scheme (obtained by exhaustive search), the greedy scheme and the random-selection scheme. From Fig.9 and Fig.10 we can see as $N$ increases, the proposed algorithm can always get the optimal throughputs. In addition, although the greedy scheme has better CU performance than the proposed algorithm (since every WLAN tends to make full use of the remaining available basic channels in the greedy scheme), the performance of the proposed algorithm is always better than the other schemes in terms of both the aggregate throughput and JFI.

In addition, we investigate when $N$ increases from 10 to 20, since the exhaustive search is infeasible in this scenario, it has been excluded from the evaluation comparison. We set $K = 17$, and the CUs of the proposed algorithm and the greedy scheme are always 1 as $N$ increases. That is, when $N$ is large, the channel allocation algorithms tend to make use of all of the available basic channels. From Fig.11, we can see the proposed algorithm always get a better throughput performance and higher JFI as $N$ increases from 10 to 20. When $N$ increases from 17 to 20, the JFI of the greedy scheme decreases while the proposed algorithm can maintain a higher JFI without throughput loss.

\begin{figure}
\begin{minipage}[t]{0.45\linewidth}
\centering
  \centerline{\includegraphics[width=1.8in]{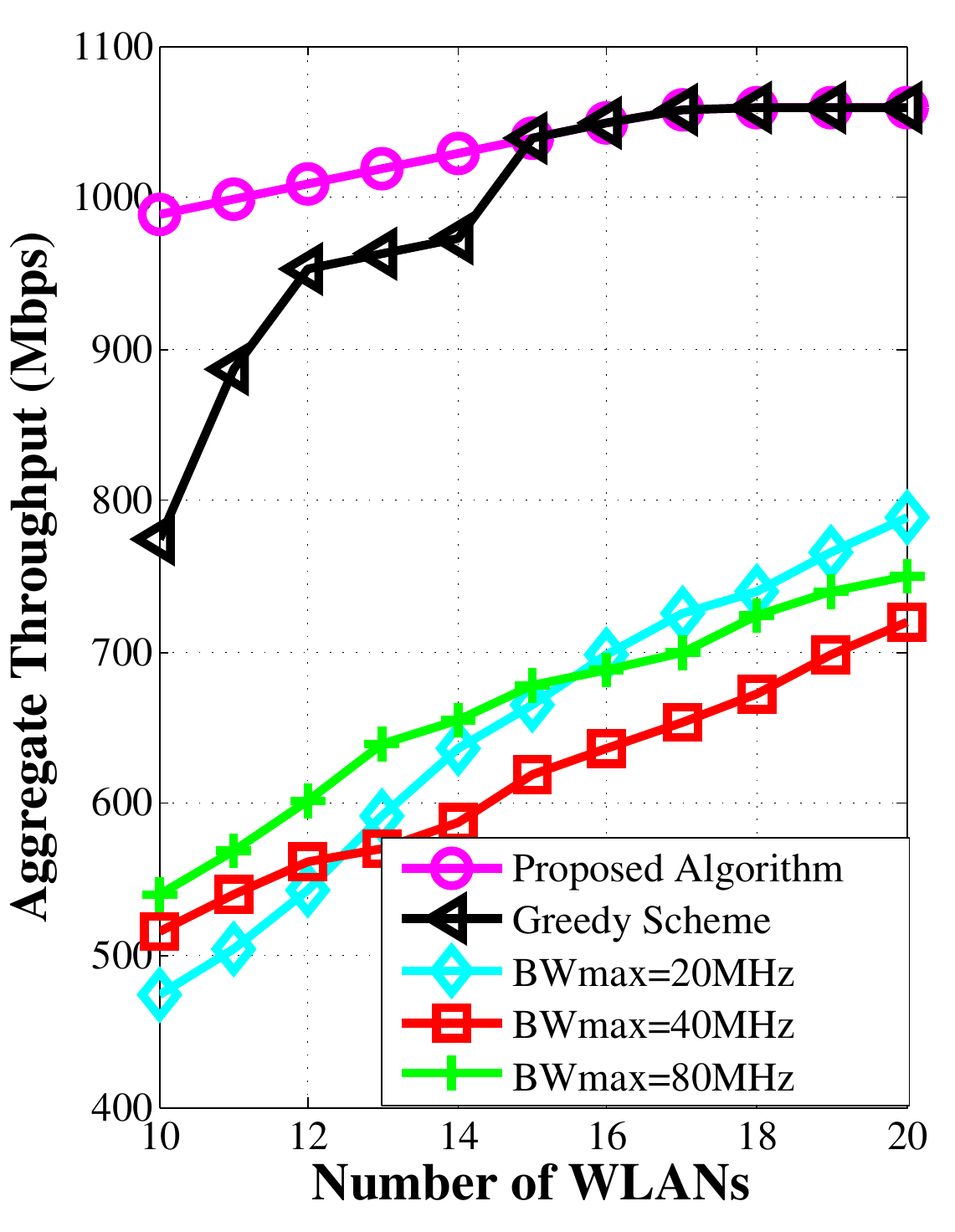}}

  \end{minipage}
  \hfill
  \begin{minipage}[t]{0.45\linewidth}
\centering
  \centerline{\includegraphics[width=1.8in]{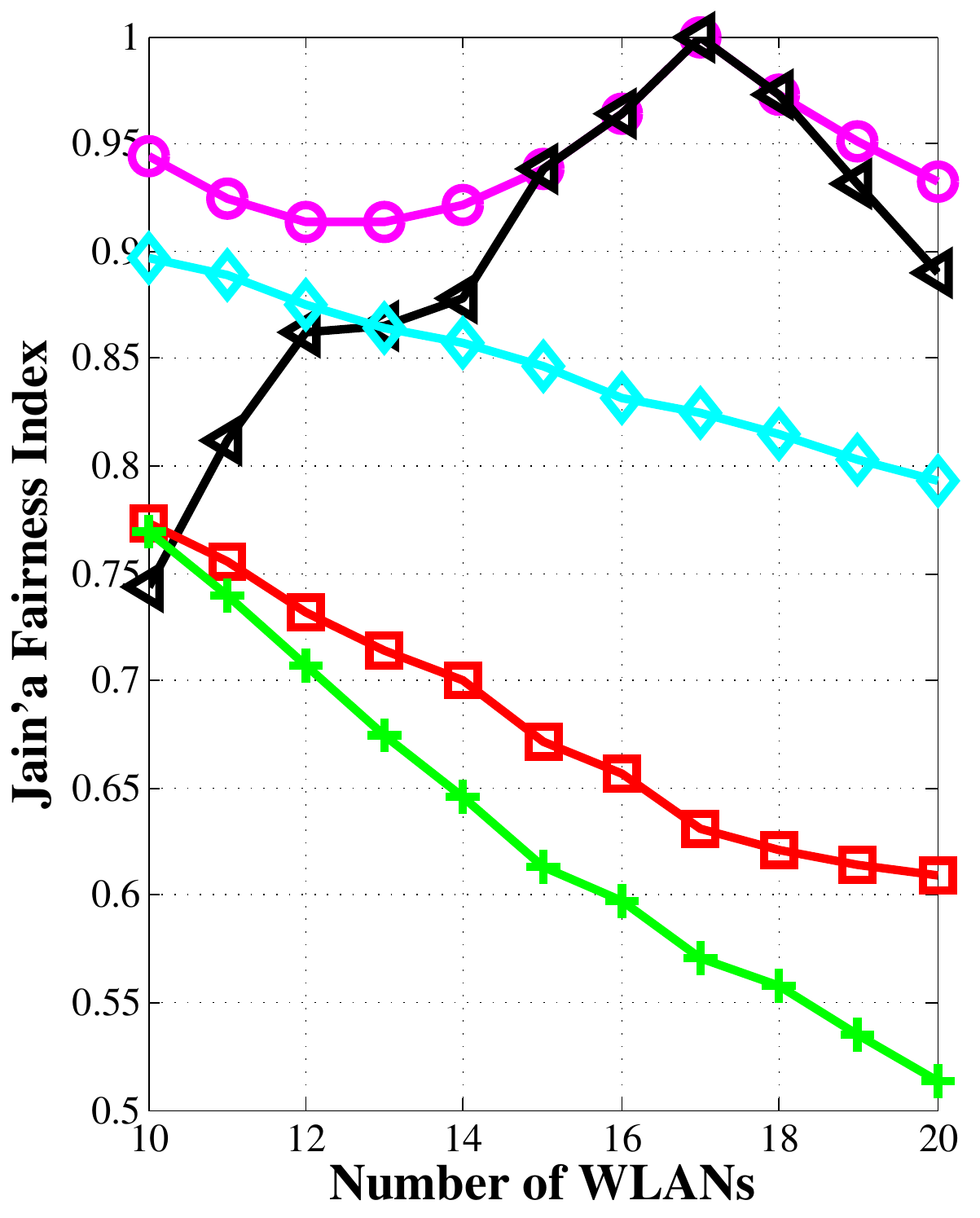}}
  \end{minipage}
  \caption{Throughput and JFI Comparison of Different Algorithms with respect to $N$ in larger size networks}
  \label{11}
\end{figure}

\section{Conclusion}
This paper investigated the channel allocation problem in “all-inclusive” DCB WLANs. Through performance analysis, we proved that the maximal throughput performance is achieved under the channel allocation scheme with the least overlapped channels. Based on this understanding, we then construct INLP models with the target of maximizing the system throughput. Based on the BBM to solve the INLP model, we proposed a channel allocation algorithm. Simulations showed that the proposed algorithm can achieve the optimal throughput performance and outperform the greedy and random-selection schemes in terms of both aggregate throughput and fairness. We believe that our analysis on the optimal channel allocation schemes brings new insights into the design and optimization of future WLANs. For instance, as pointed out, more channel bonding does not always bring more performance improvements. As a future work, we will investigate the performance of a ``non-all-inclusive" DCB network in which not all the WLANs are within the carrier-sensing range of each other and investigate efficient channel allocation algorithms to optimize system performance.

\bibliographystyle{IEEEtran}
\bibliography{journalref}

\begin{thebibliography}{10}
\providecommand{\url}[1]{#1}
\csname url@samestyle\endcsname
\providecommand{\newblock}{\relax}
\providecommand{\bibinfo}[2]{#2}
\providecommand{\BIBentrySTDinterwordspacing}{\spaceskip=0pt\relax}
\providecommand{\BIBentryALTinterwordstretchfactor}{4}
\providecommand{\BIBentryALTinterwordspacing}{\spaceskip=\fontdimen2\font plus
\BIBentryALTinterwordstretchfactor\fontdimen3\font minus
  \fontdimen4\font\relax}
\providecommand{\BIBforeignlanguage}[2]{{%
\expandafter\ifx\csname l@#1\endcsname\relax
\typeout{** WARNING: IEEEtran.bst: No hyphenation pattern has been}%
\typeout{** loaded for the language `#1'. Using the pattern for}%
\typeout{** the default language instead.}%
\else
\language=\csname l@#1\endcsname
\fi
#2}}
\providecommand{\BIBdecl}{\relax}
\BIBdecl

\bibitem{09std}
{IEEE802.11n-2009}, Standard for Wireless LAN Medium Access Control (MAC) and
  Physical Layer (PHY): Enhancements for High Throughput.

\bibitem{14std}
{IEEE802.11ac-2014}, Standard for Wireless LAN Medium Access Control (MAC) and
  Physical Layer (PHY) specifications: Enhancements for Very High Throughput
  for Operation in Bands below 6 GHz.

\bibitem{perahia2011gigabit}
E.~Perahia and M.~X. Gong, ``{Gigabit wireless LANs: an overview of IEEE 802.11
  ac and 802.11 ad},'' \emph{ACM SIGMOBILE Mobile Computing and Communications
  Review}, vol.~15, no.~3, pp. 23--33, 2011.

\bibitem{park2011ieee}
M.~Park, ``{IEEE 802.11 ac: Dynamic Bandwidth Channel Access},'' in \emph{IEEE
  ICC}, 2011, pp. 1--5.

\bibitem{gong2011channel}
M.~X. Gong, B.~Hart, L.~Xia, and R.~Want, ``{Channel Bounding and MAC
  Protection Mechanisms for 802.11 ac},'' in \emph{IEEE GLOBECOM}, 2011, pp.
  1--5.

\bibitem{deek2014intelligent}
L.~Deek, E.~Garcia-Villegas, E.~Belding, S.-J. Lee, and K.~Almeroth,
  ``{Intelligent Channel Bonding in 802.11n WLANs},'' \emph{IEEE Trans. on
  Mobile Computing}, vol.~13, no.~6, pp. 1242--1255, 2014.

\bibitem{srikanth2016performance}
S.~Srikanth and V.~Ramaiyan, ``{Performance Analysis of an IEEE 802.11ac WLAN
  with Dynamic Bandwidth Channel Access},'' in \emph{IEEE Twenty Second
  National Conference on Communication (NCC)}, 2016, pp. 1--6.

\bibitem{bellalta2016interactions}
B.~Bellalta, A.~Checco, A.~Zocca, and J.~Barcelo, ``{On the Interactions
  between Multiple Overlapping WLANs using Channel Bonding},'' \emph{IEEE
  Trans. on Vehicular Technology}, vol.~65, no.~2, pp. 796--812, 2016.

\bibitem{faridi2016analysis}
A.~Faridi, B.~Bellalta, and A.~Checco, ``{Analysis of Dynamic Channel Bonding
  in Dense Networks of WLANs},'' \emph{IEEE Trans. on Mobile Computing}, DOI:
  10.1109/TMC.2016.2615305, 2016.

\bibitem{mahonen2004automatic}
P.~Mahonen, J.~Riihijarvi, and M.~Petrova, ``{Automatic Channel Allocation for
  Small Wireless Local Area Networks using Graph Colouring Algorithm
  Approach},'' in \emph{IEEE International Symposium on Personal, Indoor and
  Mobile Radio Communications (PIMRC)}, vol.~1, 2004, pp. 536--539.

\bibitem{mishra2005weighted}
A.~Mishra, S.~Banerjee, and W.~Arbaugh, ``{Weighted Coloring based Channel
  Assignment for WLANs},'' \emph{ACM SIGMOBILE Mobile Computing and
  Communications Review}, vol.~9, no.~3, pp. 19--31, 2005.

\bibitem{chieochan2010channel}
S.~Chieochan, E.~Hossain, and J.~Diamond, ``{Channel Assignment Schemes for
  Infrastructure-based 802.11 WLANs: A survey},'' \emph{IEEE Communications
  Surveys \& Tutorials}, vol.~12, no.~1, pp. 124--136, 2010.

\bibitem{mengual2013channel}
E.~Mengual, E.~Garcia-Villegas, and R.~Vidal, ``{Channel Management in A
  Campus-wide WLAN with Partially Overlapping Channels},'' in \emph{IEEE
  International Symposium on Personal, Indoor and Mobile Radio Communications
  (PIMRC)}, 2013, pp. 2449--2453.

\bibitem{kamiya2015joint}
S.~Kamiya, K.~Nagashima, K.~Yamamoto, T.~Nishio, M.~Morikura, and T.~Sugihara,
  ``{Joint Range Adjustment and Channel Assignment for Overlap Mitigation in
  Dense WLANs},'' in \emph{Personal, Indoor, and Mobile Radio Communications
  (PIMRC), 2015 IEEE 26th Annual International Symposium on}, 2015, pp.
  1974--1979.

\bibitem{arslan2010auto}
M.~Y. Arslan, K.~Pelechrinis, I.~Broustis, S.~V. Krishnamurthy, S.~Addepalli,
  and K.~Papagiannaki, ``{Auto-configuration of 802.11n WLANs},'' in \emph{ACM
  CoNext}, 2010, p.~27.

\bibitem{gong2014distributed}
D.~Gong, M.~Zhao, and Y.~Yang, ``{Distributed Channel Assignment Algorithms for
  802.11n WLANs with Heterogeneous Clients},'' \emph{Journal of Parallel and
  Distributed Computing}, vol.~74, no.~5, pp. 2365--2379, 2014.

\bibitem{herzen2013distributed}
J.~Herzen, R.~Merz, and P.~Thiran, ``{Distributed Spectrum Assignment for Home
  WLANs},'' in \emph{IEEE INFOCOM}, 2013, pp. 1573--1581.

\bibitem{wang2016managing}
W.~Wang, F.~Zhang, and Q.~Zhang, ``{Managing Channel Bonding with Clear Channel
  Assessment in 802.11 Networks},'' in \emph{IEEE ICC}, 2016, pp. 1--6.

\bibitem{jang2015channel}
S.~Jang and S.~Bahk, ``{A Channel Allocation Algorithm for Reducing the Channel
  Sensing Reserving Asymmetry in 802.11ac Networks},'' \emph{IEEE Trans. on
  Mobile Computing}, vol.~14, no.~3, pp. 458--472, 2015.

\bibitem{gupta1985branch}
O.~K. Gupta and A.~Ravindran, ``{Branch and Bound Experiments in Convex
  Nonlinear Integer Programming},'' \emph{Management Science}, vol.~31, no.~12,
  pp. 1533--1546, 1985.

\end{thebibliography}

\end{document}